\documentstyle[pre,aps,preprint,tighten,eqsecnum]{revtex}
\begin{document}
\draft
\title{ DYNAMICS OF A SIMPLE QUANTUM SYSTEM IN A COMPLEX ENVIRONMENT }
\author{Aurel BULGAC \cite{AB} }
\address{ Department of Physics, University of Washington,\\
P.O. Box 351560, Seattle, WA 98195--1560, USA}
\author{ Giu DO DANG \cite{GDD}}
\address{ Laboratoire de Physique Th\'{e}orique et Hautes Energies,\\
Universit\'{e} de Paris--Sud, B\^{a}t. 211, 91405 Orsay, FRANCE }
\author{ Dimitri KUSNEZOV \cite{DK}}
\address{ Center for Theoretical Physics, Sloane Physics Laboratory, \\
Yale University, New Haven, CT 06520--8120, USA}
\date{\today}
\maketitle
\begin{abstract}
We present a theory for the dynamical evolution of a quantum system
coupled to a complex many--body intrinsic system/environment. By
modelling the intrinsic many--body system with parametric random
matrices, we study the types of effective stochastic models which emerge
from random matrix theory. Using the Feynman--Vernon path integral
formalism, we derive the influence functional and obtain either
analytical or numerical solutions for the time evolution of the entire
quantum system. We discuss thoroughly the structure of the solutions for
some representative cases and make connections to well known limiting
results, particularly to Brownian motion, Kramers classical
limit and the Caldeira--Leggett approach.
\end{abstract}

\vspace{0.5cm}


\vspace{1.5cm}

\newpage



\section{ INTRODUCTION }

Quantum dissipation is a problem with such a long history and such a
multitude of results that even a cursory review will not make justice to
the numerous contributions of a large number of authors over several
decades. We recommend the reader to consult Refs.
\cite{feyn,cald,weiss,diosi,brink,bdk_ann,bdk_mb8,bdk_pre,chaos,cohen}
and the references therein. In spite of this impressive effort,  many
people do  not consider the problem of quantum dissipation  a solved
issue, and its character and microscopic origin still call for the
attention of a large community across many, if not all, subfields of
physics. The present paper represents a continuation of our effort to
understand the character of energy flow between the slow degrees of
freedom and the intrinsic degrees of freedom in many--body systems. Our
initial motivation was to understand the ``irreversible'' time evolution
of the large amplitude nuclear collective motion
\cite{bdk_ann,bdk_mb8,bdk_pre,chaos,bdk_pla} and
for that reason we have adopted
a traditionally nuclear physics approach. The large body of evidence,
both experimental and theoretical, suggests that many Fermion systems
can be described reasonably well within the framework of a random matrix
formalism. Another way of saying the same thing is that a many Fermion
system is predominantly a quantum chaotic system and thus a random
matrix approach is a natural approach. At the same time, both theory and
experiment strongly suggest that there are some degrees of freedom,
which are not chaotic and usually are referred to as collective or
shape. However these collective degrees of freedom are coupled with a
large number of non--collective degrees of freedom and as a result a
rather generic situation results: a relatively small quantum system in
contact with an ``environment''. Even though the whole system is finite
and in a strict sense there is no irreversible behaviour in this case,
for all practical purposes the time evolution of the collective or slow
quantum system has the character of quantum dissipative dynamics. In our
formulation, we will neglect other physical mechanisms which lead to
dissipative contributions, such as particle evaporation or coupling to
electromagnetic fields. This can be viewed as limiting the present
study to shorter time scales. Such phenomena can in principle be
introduced into the formalism, but we shall not attempt it here. We
shall not either try here to defend the legitimacy of such a terminology
(dissipation), the issue at stake however is unquestionably sound. The
reader will  recognize easily that the problem we address here is
typical and under different guises appears in many subfields of physics.

As an introduction, let us consider for the moment that a certain simple
system interacts with some relatively large (but finite) many--body
system. The question is: can one describe the dynamical behaviour of
the simple system using for example an equation of the form
\begin{equation}
M\frac{d^2X}{dt^2} =-\frac{d U(X)}{dX} -
 M\gamma \frac{dX}{dt}+f(X,t) \label{eom}
\end{equation}
as in the case of a Brownian particle, if in the absence of the
interaction the Hamiltonian of the system is
\begin{equation}
 H_0(X,P)=\frac{P^2}{2M}+ U(X) \label{h0}
\end{equation}
and where $\gamma $ is a friction coefficient and $f(X,t)$ is a
Langevin--like force? The force $f(X,t)$ can in principle depend not
only on time but on position as well, and in this way one can describe a
large variety of physical situations, ranging from diffusion to
localization \cite{cohen}. If one were to start from a description of
the entire system (reservoir plus simple system) with a Hamiltonian
\begin{equation}
H(X,P,x,p) = H_0(X,P) + H_1(X,x,p) \label{toth} ,
\end{equation}
where $ H_1(X,x,p)$ describes the reservoir and its interaction with our
system, under what circumstances could one derive an equation of motion
like Eq.(\ref{eom})? Moreover, does the fluctuating force have Gaussian
character or not? Since our approach is a fully quantum mechanical one,
we shall be able to answer another important question, on the specific
role played by the quantum effects. As we shall show, the classical
results are recovered when the appropriate limits are taken, however,
the classical limit is not at all a trivial one nor is it reached in a
simple fashion. If one were to take the approach of postulating that an
equation of the type Eq. (\ref{eom}) governs the dynamics and assume for
example that the fluctuation properties of $f(X,t)$ are Gaussian in
character, then the entire powerful apparatus developed for Brownian
motion could be then invoked \cite{cohen}. But such an approach would leave
unanswered the main questions of whether one can describe in this way
collective degrees of freedom in a finite closed system, like atomic
nuclei.

We address this problem using a well known approach  based on the double
path integral formulation of Feynman and Vernon \cite{feyn}. Our
original input is in the functional form of the influence functional,
which arises from a parametric random matrix description of the
``environment''. This has been attempted earlier in Ref. \cite{brink}.
The functional form for the influence functional we determine is
qualitatively different from the popular Caldeira--Leggett type derived
by Feynman and Vernon \cite{feyn}.  The parameters which define the
influence functional have a rather simple and intuitive meaning from a
microscopic point of view and we refer the reader to earlier
publications for details and discussions
\cite{brink,bdk_ann,bdk_mb8,bdk_pre,chaos,bdk_pla}.
It comes as no surprise than
that under such circumstances the dynamical evolution of a quantum
dissipative system in our case has new features as well, as we shall
amply exemplify in the body of the paper. Here we restrict our attention
to the Markovian limit only and we hope to address the important problem
of memory effects in the future. In spite of its physical restrictions
(high temperature limit for the intrinsic system) this limit shows
already the qualitative differences with the previously known
approaches.

The paper is organized as follows. In Section II we discuss the time
evolution equation for the density matrix of the ``slow'' quantum system
coupled to a complex many--body system. In Section III we show that, at
high temperatures, the evolution equation for the density matrix can be
brought to the Kramers form, when the classical limit is taken. Sections
IV--VII discuss exact solutions to the evolution equations for certain
potentials. Section VIII is devoted to the study of the tunneling
problem. A short summary and discussion of the results is given in the
final Section.

\section{ EVOLUTION EQUATION FOR THE DENSITY MATRIX }

In this section we discuss the description of the internal degrees of
freedom (or complex environment) through parametric random matrix
theory, and derive the equation of motion for the density matrix of the
slow degrees of freedom by integrating over the internal states.

\subsection{ Random Matrix Model}

The basic assumption concerning the intrinsic states is that there are
no governing constants of the motion, so that the dynamics is chaotic.
This has been seen to be the general situation in studies of many--body
systems, from nuclei to molecules, so it is reasonable to approach the
modelling of these degrees of freedom with random matrices, suitably
tailored to the problem. The generic form of the Hamiltonian governing
the dynamics of a ``slow'' quantum system coupled to a complex
environment is described as follows
\begin{equation}
H(X,x) = H_0(X) + H_1(X,x) =
\frac{P_X^2}{2M} +U(X) + H_1(X,x)\label{totham}.
\end{equation}
We shall often refer to $X$ as ``shape'' variables, since in large
amplitude collective nuclear motion it represents the collective
coordinates which characterize the nuclear mean field.

The part of the total Hamiltonian Eq. (\ref{totham}) which depends on
the intrinsic coordinates $H_1(X,x)$ is defined as a matrix, whose
matrix elements depend parametrically on the ``slow'' coordinate $X$
\begin{equation}
[H_1(X)]_{ij} = [h_{0}]_{ij} + [h_{1}(X)]_{ij}.
\end{equation}
$h_0$ is taken to be diagonal and defines the average density of states,
with $\langle k|h_0|l\rangle = [h_0]_{kl}=\varepsilon _k \delta _{kl}$.
We refer in the main text to these eigenstates as ``typical states'' of
the intrinsic system with an energy $\varepsilon$. One can think of
$h_0$ as a Hamiltonian describing a ``bath'' or a ``reservoir'' and of
$h_1(X)$ as a Hamiltonian describing the interaction between the
``bath/reservoir'' and the ``slow'' system. Whereas in statistical
physics the interaction between the thermostat and the system under
consideration is assumed to be negligible, we shall not make such an
approximation here. As a matter of fact, for the physically interesting
situations we envision, this coupling term can be large. This fact alone
leads to significant differences of various distributions when compared
with the corresponding results of traditional approaches.

For an intrinsic subsystem with a large number of degrees of freedom,
the average density of states,
\begin{equation}
\rho (\varepsilon ) =
\overline{\mathrm{Tr} \delta (H_1(X)-\varepsilon )},
\end{equation}
for each given shape $X$ increases sharply with energy. The overline
denotes here a procedure for extracting the smooth part of $\rho
(\varepsilon )$ as a function of energy, which amounts essentially to an
ensemble average, to be introduced below. For a many Fermion system,
$\rho (\varepsilon )$ has a roughly exponential behaviour. Recall that
$\ln \rho (\varepsilon )$ is approximately proportional to the
thermodynamic entropy of the intrinsic system, which is an extensive
quantity. The fact that the average density of states for the intrinsic
subsystem has such a behavior is a key element of the entire approach.
This is equivalent to stating that the intrinsic subsystem has a large
heat capacity and thus can play the role of a ``reservoir'', although
not necessarily ideal. In principle $\rho (\varepsilon )$ can be
$X$--dependent as well, but we shall ignore this aspect here. Without an
$X$--dependence of the average density of states, mechanical work cannot
be performed on or by the model environment we study here, and only heat
exchange is allowed.

In Refs. \cite{bdk_ann,bdk_mb8,bdk_pre,chaos,bdk_pla} we discussed the reasons
why one chooses this specific form  of the Hamiltonian. In the basis of
the eigenstates of $h_0$, we define $h_1(X)$ as a parameter dependent,
$N\times N$ real Gaussian random matrix, which is completely specified
by its first two moments
\begin{eqnarray}
\overline{[h_1(X)]_{kl}} &=&0,\nonumber\\
\overline{ [h_1(X)]_{ij}[h_1(Y)]_{kl} } &=&
[\delta _{ik}\delta _{jl}+\delta _{il}\delta_{jk}]{\cal G}_{ij}(X-Y).
\label{correl}
\end{eqnarray}
The overline stands for statistical averages over the ensemble of random
Gaussian matrices from the Gaussian Orthogonal Ensemble (GOE)
\cite{mehta}. ${\cal G}_{ij}(X-Y)$ can be taken as a ``bell shaped''
correlation function with a characteristic width $X_0$, or, in some
physically interesting cases  even periodic, with period $\propto X_0$.
The fact that this correlator is ``translational invariant'' is not a
crucial limitation, and a general form can be adopted without any
significant changes in the formalism. We limit our analysis to the GOE
case only for the sake of simplicity of the argument, as any other
Gaussian ensemble can be treated in a similar manner. The dependence on
$i,j$ allows for the description of banded matrices, where an effective
number of states $N_0\leq N$ are coupled by $h_1(X)$. It is convenient
to use an explicit parameterization, which incorporates the average
density of states and the bandwidth of the statistical fluctuations
explicitly \cite{brink}:
\begin{equation}
{\cal G}_{ij}(X)=
\frac{\Gamma ^\downarrow }{2\pi \sqrt{\rho (\varepsilon _i)\rho (\varepsilon
_j)}}
\exp \left [ -\frac{(\varepsilon _i -\varepsilon _j)^2}{2\kappa _0
^2} \right ]G \left (\frac{X}{X_0}\right ).
\end{equation}
Here $G(x)=G(-x)=G^*(x)\le 1$, $G(0)=1$, and the spreading width $\Gamma
^\downarrow$, $\kappa_0$ (linked with the effective band width $N_0
\approx \kappa _0 \rho(\varepsilon ))$ and $X_0$ are characteristic of
the intrinsic system. Even though it is not necessary, in this paper we
shall consider a particular from for $G(x)$, namely $G(x)=\exp(-x^2/2)$.

To complete the tailoring of the random matrices, we require a realistic
average density of states for the reservoir. It is reasonable to assume
that in a suitable energy interval, $\rho$  has the behaviour
\cite{brink}
\begin{equation}
 \rho(\varepsilon )=\rho _0 \exp (\beta \varepsilon),\qquad
 \beta = \frac{1}{T}= \frac{d}{d \varepsilon} \ln \rho (\varepsilon ).
\end{equation}
$\beta $ can thus be interpreted as the thermodynamic temperature of the
intrinsic system. As $\beta$ is independent of the internal excitation
energy $\varepsilon $, this particular type of intrinsic quantum system
plays the role of a ``perfect'' thermal quantum reservoir. We note in
passing that even though the temperature of the ``reservoir'' remains
constant throughout the entire dynamical evolution of the whole system,
one should not conclude from this that the ``reservoir'' is in thermal
equilibrium. As we have shown explicitly in Refs. \cite{bdk_pre,chaos}
the population of various energy levels of a uniformly driven
``reservoir'' is far from an equilibrium Boltzmann distribution. The
parameters of the present construction are summarized in Table I.

\subsection{ Influence Functional}

The quantum description of our coupled system will be treated through
the path integral construction of the density matrix. According to
Feynman and Vernon \cite{feyn}, one can write the following double path
integral representation for the density matrix of the entire system
\begin{eqnarray}
{\cal{R}}(X,x,Y,y,t) & = & \int d X_0 dY_0
\psi (X_0) \psi ^*(Y_0)
\int _{X(0)=X_0} ^{X(t)=X} {\cal{D}}X(t)
\int _{Y(0)=Y_0} ^{Y(t)=Y} {\cal{D}}Y(t) \nonumber \\
& \times & \exp \left \{ \frac{i}{\hbar }
\left [ S_0(X(t)) - S_0(Y (t)) \right ] \right \} \nonumber \\
 & \times & \langle x |{\mathrm{T}} \exp \left [
-\frac{i}{\hbar}
\int _{0}^{t} dt^\prime H_1(X(t^\prime )) \right ] | \phi \rangle
\langle \phi | {\mathrm{T}}_a \exp \left [
 \frac{i}{\hbar}
\int _{0}^{t} dt^{\prime \prime} H_1(Y(t^{\prime \prime }))
\right ] | y \rangle ,
\end{eqnarray}
where ${\mathrm{T}}$ and ${\mathrm{T}}_a$ represent the time ordering and
time anti--ordering operators respectively. In this representation, we
have used a particular form for the initial state wave function,
\begin{equation}
\Psi (X,x)=\psi (X)\phi (x).
\end{equation}
Other choices are equally possible, such as an initial density matrix.
By introducing the influence functional ${\cal{L}}(X(t),Y(t),t)$
\begin{equation}
{\cal{L}}(X(t),Y(t),t)=
\langle \phi | \left \{ \! {\mathrm{T}}_a \! \exp \left [
 \frac{i}{\hbar} \int _{0}^{t} dt^{\prime \prime }
H_1(Y(t^{\prime \prime })) \right ] \right \} \!
\left \{ \! {\mathrm{T}} \! \exp \left [
-\frac{i}{\hbar} \int _{0}^{t} dt ^\prime H_1(X(t^\prime ))
\right ] \right \}
| \phi \rangle
\end{equation}
one readily obtains the following double path integral representation
for the density matrix for the ``slow'' subsystem
\begin{eqnarray}
\rho (X,Y,t) & = & \int d X_0 dY_0
\psi (X_0) \psi ^*(Y_0)
\int _{X(0)=X_0} ^{X(t)=X} {\cal{D}}X(t)
\int _{Y(0)=Y_0} ^{Y(t)=Y} {\cal{D}}Y(t) \nonumber \\
 & \times & \exp \left \{ \frac{i}{\hbar }
\left [ S_0(X(t)) - S_0(Y (t)) \right ] \right \} {\cal{L}}(X(t),Y(t),t).
\end{eqnarray}
The formulation of the problem through a path integral representation
serves only as a very convenient vehicle to obtain an evolution equation
for the density matrix $\rho (X,Y,t)$.

\subsection{ Evolution Equations}

The evolution equation for the influence functional
${\cal{L}}(X(t),Y(t),t)$  has been solved in Ref. \cite{bdk_pre,chaos,bdk_pla}
for the case $N\rightarrow \infty$ and  the case when the
``temperature'' of the reservoir is infinite. In Appendices A--D we
compute the first order correction in $\beta$ to the influence
functional in the adiabatic limit, when the characteristic time scale of
the ``reservoir'' $\hbar/\kappa _0$ is significantly shorter than the
characteristic time scale of the slow system for which we derive the
dynamical evolution equation. We thus obtain for the influence
functional
\begin{eqnarray}
{\cal{L}}(X(t),Y(t)) &=&
\exp \left \{
\frac{\Gamma ^\downarrow }{\hbar } \int _0^ t ds \left [ G\left (
\frac{X(s)-Y(s)}{X_0} \right )- 1 \right ] \right \} \nonumber \\
&\times & \exp \left \{   \frac{i\beta \Gamma ^\downarrow }{4X_0}
\int _0^ t ds [
\dot{X}(s)+\dot{Y}(s)] G^\prime \left ( \frac{X(s)-Y(s)}{X_0}\right )
\right \} . \label{infl}
\end{eqnarray}
where $G^\prime (x)=dG(x)/dx$. The physical significance of all other
quantities entering this expression has been explained and discussed at
length in Refs. \cite {bdk_ann,bdk_mb8,bdk_pre,chaos}, and is briefly
summarized in Table I.  It is worth noting that the functional form of
the influence functional derived by us is different from the
Caldeira--Leggett form \cite{cald}, which is a quadratic expression in
$X(t)$ and $Y(t)$. If we were to use only the first term in a Taylor
expansion of $G[(X(s)-Y(s))/X_0]-1$, we would obtain an expression
similar to Caldeira--Leggett form for the influence functional.
However, the present form of the influence functional leads
in the classical limit to a velocity dependent frictional force, see
Ref. \cite{bdk_pre,chaos}.

By combining the double path integral representation for the density
matrix $\rho (X,Y,t)$ with the above expression for the influence
functional in the adiabatic approximation one readily obtains that the
density matrix satisfies the following Schr\"{o}dinger like equation
(for similar examples see Refs. \cite{cald,diosi})
\begin{eqnarray}
i\hbar \partial _t \rho (X,Y,t) &=&
\left \{ \frac{P_X^2}{2M} - \frac{P_Y^2}{2M} + U(X) - U(Y) \right .
\label{evol}\\
 &-& \left .
\frac{ \beta \Gamma ^\downarrow \hbar }{4X_0M}
G^\prime \left ( \frac{X-Y}{X_0}\right ) (P_X-P_Y)
+  i \Gamma ^\downarrow
\left [ G \left ( \frac{X-Y}{X_0}\right )-1\right ] \right \} \rho (X,Y,t)
 \nonumber
\end{eqnarray}
with an arbitrary initial condition
\begin{equation}
\rho (X,Y,0) = \rho _0(X,Y).
\end{equation}
This equation is the central object of our study and the remaining of
the paper is devoted to determining various limiting regimes and the
character of its solutions upon varying $\beta $, $\Gamma ^\downarrow $,
$X_0$ and $\hbar $. (In this paper we have already taken the limit
$\kappa _0 \rightarrow \infty$.) At first glance the reader might get
the impression that the slow subsystem we consider here is characterized
by one degree of freedom only. As a simple analysis will show however
that the slow subsystem can have an arbitrary number of degrees of
freedom and most of the formulas we shall present are equally valid in
this case.

This evolution equation is somewhat peculiar in certain aspects. It is
obvious that in the absence of the coupling to the ``reservoir'', it
describes a purely quantum evolution of the ``slow'' subsystem. Eq.
(\ref{evol}) has been derived from a purely quantum description of the
entire system, by performing the expansion in $\beta$. In Eq.
(\ref{evol}) however, the inverse temperature enters only in the
combination $\tau _{th}=\beta \hbar$, which can be interpreted as a
thermal time (analogous to the thermal de Broglie wave length). Thus the
expansion in the inverse temperature $\beta$ is at the same time an
expansion in $\hbar$. Our limitation to the zeroth and first order
terms in $\beta\hbar$  for the coupling between the ``slow'' subsystem
and the ``reservoir'' can consequently be interpreted as a semiclassical
approximation.

It has been argued by Diosi \cite{diosi} that for the case of a
Caldeira--Leggett correlator ($G(x)=1-x^2/2$) the similar high
temperature limit of the evolution equation requires the retention of
the next order term in $\beta$ in order to bring the corresponding
approximate evolution equation to a Lindblad form \cite{lind}, which
maintains positivity of the density distribution for any physically
acceptable initial conditions. If these higher order terms in $\beta$
are not introduced, such an equation (as Eq. (\ref{evol})) cannot be
applied to an initial state which is narrower than the thermal de
Broglie wave length, $\lambda_T=2\pi\hbar\sqrt{\beta/M}$. This
restriction to wave packets which are wider than the thermal de Broglie
wave length is manifest in a somewhat different way as well. Let us
compute the rate of change of the ``total energy'' of the ``slow''
subsystem, which can be defined naturally  as follows:
\begin{equation}
E_0(t)=\mathrm{Tr} [H_0\rho (t)].
\end{equation}
Using the evolution equation Eq. (\ref{evol}) for the density matrix
$\rho (X,Y,t)$, after some straightforward manipulations one obtains the
following expression for the rate of change of the ``total energy'':
\begin{equation}
\frac{dE_0(t)}{dt}=
2 \gamma \left [ \frac{T}{2} -\frac{\langle P^2\rangle }{2M} \right ],
\label{totener}
\end{equation}
where $\gamma$ is the friction coefficient in the small velocity limit
to be introduced below, see Eq. (\ref{frict}). This rate has an
apparently pure classical content. This is of course deceiving, as
quantum effects are clearly retained in  both Eqs. (\ref{evol}) and
(\ref{totener}),  even though not entirely in Eq. (\ref{totener}).
However, since one has to assume that $T>\hbar^2/ML^2$, where $L$ is the
characteristic spatial extension of the state, it is possible in Eq.
(2.15)  to replace the quantity $\langle P^2\rangle =\langle P\rangle ^2
+\langle \!\langle P^2\rangle \! \rangle $ with simply $\langle P\rangle
^2$ . This renders Eq. (2.15) purely classical in character. One incurs
a certain loss of accuracy and a small degree of inconsistency by
proceeding in this manner, so it is better to leave Eq. (\ref{totener})
as is.

Even though one can go beyond the first order in $\beta$ and one can
derive a more accurate evolution equation for the density matrix $\rho
(X,Y,t)$ in the high temperature limit, we shall not do that in this
work, for the sake of simplicity of the presentation. We do not expect
that such corrections will lead to a qualitatively new behaviour.

\subsection{Coordinate and Momentum Distributions and the Cumulants
Expansion}

We will derive the time dependent solutions $\rho(X,Y,t)$ in the
following sections, and from that it will be useful to extract
information concerning the behavior of coordinates and momenta. The most
convenient way to do so is through the cumulant expansion. To define
this, we start by introducing the new variables:
\begin{equation}
r=\frac{X+Y}{2}, \;\;\; s=X-Y \label{var} .
\end{equation}
Coordinate and momentum information can readily be extracted from the
following rather atypical Fourier transform of the density matrix:
\begin{equation}
\rho (r,s,t)= \int \frac{d k}{2\pi \hbar}
\exp \left ( \frac{ikr}{\hbar} \right ) d(s,k,t)
\label{dtran}.
\end{equation}
For either $s=0$ or $k=0$, $d(s,k,t)$ is the characteristic function
\cite{vankampen} for the spatial or momentum distribution of a given
quantum state respectively.  For example, if we are interested in the
spatial diffusion $X$, as measured by $\langle\!\langle
X^2\rangle\!\rangle$, then we get from Eq. (\ref{dtran}) and integration
by parts:
\begin{equation}
\langle X^2\rangle = \int dX' \rho(X',X',t) {X'}^2
     =-\hbar^2\frac{d^2}{dk^2} d(0,k,t)\mid_{k=0}
      =\langle \!\langle X^2 \rangle \!\rangle
       +\langle X\rangle ^2.
\end{equation}
Similarly, in order to compute the average collective energy one needs
$\langle P^2\rangle$:
\begin{eqnarray}
\langle P^2\rangle &=& \int dX dX' \rho(X,X',t)
       \langle X'\mid P^2\mid X\rangle\\\nonumber
     &=&\frac{1}{2\pi\hbar}\int dX dX' dP\rho(X,X',t)
    \exp \left [ \frac{iP(X'-X)}{\hbar}\right ]
        P^2\\ \nonumber
     &=&\int ds dP  \exp \left ( \frac{iPs}{\hbar}\right )
     d(s,0,t) P^2 =
  - \hbar^2\frac{d^2}{ds^2} d(s,0,t)\mid_{s=0}
      =\langle \!\langle P^2 \rangle \!\rangle
        +\langle P\rangle ^2.
\end{eqnarray}
The quantities denoted $\langle \!\langle \cdots \rangle \!\rangle$ are
the cumulants of the distribution.

From the definition of the function $d(s,k,t)$ one defines the general
coordinate and momentum cumulant expansion as:
\begin{eqnarray}
\ln d(s,k,t)|_{s=0} &=&\sum _{n=1}^{\infty } \frac{1}{n!}\left (
\frac{ik}{\hbar } \right ) ^n
\langle \!\langle r^n \rangle \!\rangle \\
\ln d(s,k,t)|_{k=0} &=& \sum _{n=1}^{\infty } \frac{1}{n!}\left (
\frac{s}{i\hbar } \right ) ^n
\langle \!\langle p^n \rangle \!\rangle
\end{eqnarray}
where $\langle \!\langle r^n \rangle \!\rangle $ and $\langle \!\langle
p^n \rangle \!\rangle$ are the (time dependent) cumulants of the spatial
and momentum distribution respectively. One can show that the zeroth
order terms in both cumulant expansions vanish, which is consistent with
the fact that the probability is conserved within the present formalism.

A Gaussian process has only non--vanishing first and second cumulants.
In general it is known from Marcienkiewics'\cite{marc} or Pawula's
\cite{vankampen} theorem, that for a probability distribution, one
either has a Gaussian process with only the first two cumulants
non--vanishing, or all cumulants are present. Furthermore, while there
are some inequalities which relate cumulants of varying order, in most
cases there is no restriction on their sign, which can be positive or
negative.

\subsection{ Intrinsic States}

By ``integrating'' over the intrinsic variables one of course looses a
lot of information and only some average behavior of the slow subsystem
can be thus inferred. In principle one can recover some of this
information in the following manner. As in Refs.
\cite{bdk_ann,bdk_mb8,bdk_pre,chaos} one can introduce instead the
generalized occupation number probabilities
\begin{eqnarray}
{\cal {N}}(X(t),Y(t),t,\varepsilon )& = &
\langle \phi | {\mathrm T}_a \exp \left [
 \frac{i}{\hbar} \int _{0}^{t} dt^{\prime \prime }
H_1(Y(t^{\prime \prime })) \right ]
|\varepsilon \rangle \nonumber \\
& \times & \langle \varepsilon | {\mathrm T} \exp \left [
-\frac{i}{\hbar} \int _{0}^{t} dt ^\prime H_1(X(t^\prime ))
\right ]
| \phi \rangle
\end{eqnarray}
where $|\varepsilon \rangle $ is a ``typical'' state of the intrinsic
subsystem with energy $\varepsilon $. One can derive now evolution
equations for these quantities. With the help of these generalized
occupation number probabilities one can introduce what was called in
Refs. \cite{bdk_pre,chaos} the characteristic functional
\begin{equation}
{\cal {N}}(X(t),Y(t),t,\tau ) =
\int d\varepsilon \rho (\varepsilon
){\cal {N}}(X(t),Y(t),t,\varepsilon )
\exp \left ( \frac{i\varepsilon \tau }{\hbar }
 \right ) ,
\end{equation}
where $\rho (\varepsilon )$ is the average level density for the
intrinsic subsystem.  From this expression, by performing an inverse
Fourier transform one can reconstruct the generalized occupation
probabilities ${\cal {N}}(X(t),Y(t),t,\varepsilon )$. In such a case
there is a double path integral representation for the density matrix
for the slow subsystem with the condition that the intrinsic subsystem
has a well defined energy $\varepsilon $, namely
\begin{eqnarray}
\rho (X,Y,t,\varepsilon ) & = & \int d X_0 dY_0
\psi (X_0) \psi ^*(Y_0)
\int _{X(0)=X_0} ^{X(t)=X} {\cal{D}}X(t)
\int _{Y(0)=Y_0} ^{Y(t)=Y} {\cal{D}}Y(t) \nonumber \\
& \times & \exp \left \{ \frac{i}{\hbar }
\left [ S_0(X(t)) - S_0(Y (t)) \right ] \right \}
\rho (\varepsilon ) {\cal{N}}(X(t),Y(t),t,\varepsilon ).
\end{eqnarray}
In principle this representation can be used to extract more detailed
information concerning the time evolution of the coupled slow and
complex intrinsic subsystems. We shall not attempt to do that however in
this paper.

\section{ CLASSICAL LIMIT: Kramers Equation}

It is interesting to explore the classical transport equation which
emerges from Eq. (\ref{evol}). The standard approach is to construct
the Wigner transform $f(Q,P,t)$ of the density matrix $\rho(X,Y,t)$ as
\begin{equation}
  f(Q,P,t) = \frac{1}{2\pi\hbar}\int dR
   \exp \left (- \frac{iPR}{\hbar} \right )
  \rho\left ( Q+\frac{1}{2}R,Q-\frac{1}{2}R,t\right ) .
\end{equation}
It is well known that while  $f(Q,P,t)$ can be interpreted as a
classical probability distribution in a phase space $(Q,P)$, it is only
a quasi--probability since its sign can be positive or negative at a
given phase space point, while its integral over any unit phase space
cell (of size $\hbar$) is positive semi--definite. We will further
introduce the friction coefficient:
\begin{equation}
\gamma =\frac{\beta \Gamma ^\downarrow \hbar }{2MX_0^2}. \label{frict}
\end{equation}
As we will see, this definition will emerge naturally from our analysis
of the dynamical evolution of the quantum systems. However, in taking
the Wigner transform of $\rho(X,Y,t)$, we must also take the classical
limits of the quantities $\Gamma ^\downarrow$ and $X_0$, which have an
intrinsic quantum interpretation. While it is not entirely clear to us
how to define such limits, we will see that the combination
which appears in $\gamma$ has a natural classical interpretation.

From the definition, one can readily compute the classical evolution
equation, which is simply the Wigner transform of the right hand side of
(\ref{evol}):
\begin{eqnarray}
  \partial_t f(Q,P,t) &=&
  \frac{1}{2\pi\hbar}\int dR
   \exp \left (- \frac{iPR}{\hbar} \right )
   \partial_t \rho\left ( Q+\frac{1}{2}R,Q-\frac{1}{2}R,t\right ) \\
 &=&  \frac{1}{2i\pi\hbar ^2}\int dR
   \exp \left (- \frac{iPR}{\hbar} \right )
 \left\{-\frac{\hbar^2}{2M}\partial_Q\partial_R +
        U\left ( Q+\frac{1}{2}R\right )-
        U\left ( Q-\frac{1}{2}R\right ) \right.\nonumber \\
 & & \left.+i\Gamma^\downarrow\left [1-G\left (
         \frac{R}{X_0}\right)\right ]
      -i\gamma\hbar X_0G'\left (\frac{R}{X_0}\right )\partial_R\right\}
      \rho \left ( Q+\frac{1}{2}R,Q-\frac{1}{2}R,t \right ) \nonumber
\end{eqnarray}
For the contribution due to the potential energy $U(X)$, one can use the
Kramers--Moyal expansion
\begin{eqnarray}
  \int \frac{dR}{2\pi\hbar} & &
\exp \left ( -\frac{iPR}{\hbar}\right )
\left [ U\left ( Q+\frac{1}{2}R\right ) -
        U\left ( Q-\frac{1}{2}R\right ) \right ]
  \rho\left ( Q+\frac{1}{2}R,Q-\frac{1}{2}R,t \right ) \\ \nonumber
 & =& U(Q) \frac{2}{\hbar}
  \sin\left ( \frac{\hbar
\loarrow{\partial}_Q\roarrow{\partial}_P}{2} \right )  f(Q,P,t)\\
 &=&\partial_Q U(Q) \partial_P f(Q,P,t) + o(\hbar)\nonumber
\end{eqnarray}
where $\loarrow{\partial}_Q$ and $\roarrow{\partial}_P$ in the sine term
act only on $U$ and $f$, respectively. In the last line only
terms to $o(\hbar)$ were retained. For the terms which depend upon the
correlation function $G(x)$, we consider a general expansion
\begin{equation}
G(x)\approx 1-\frac{x^2}{2} + ...
\end{equation}
The terms beyond the quadratic ones  in the expansion will come with
higher powers of $\hbar$ into the evolution equation and are hence
omitted.  Integrating by parts and using the fact that $\rho(X,Y,t)$
vanishes when $X,Y\rightarrow\pm\infty$, we have:
\begin{eqnarray}
\frac{ \partial f (Q,V,t)}{\partial t} &+&
V\frac{\partial  f (Q,V,t)}{\partial Q}
-\frac{1}{M}
\frac{\partial U(Q) }{\partial Q}\frac{\partial  f (Q,V,t)}{\partial
V}\\
  & = &\gamma \left \{ \frac{\partial [ Vf(Q,V,t)]}{\partial V}+
  \frac{T}{M} \frac{\partial ^2f (Q,V,t)}{\partial V^2} \right \},\nonumber
\end{eqnarray}
where $T=1/\beta $ is the temperature, and the velocity is
$V=\dot{Q}=P/M$. We have thus obtained Kramers equation \cite{vankampen}.

We will derive transport coefficients below, and it is worth noting that taking
classical limits is not straightforward. In quantum Brownian motion, we will
extract a diffusion constant which is related to the friction $\gamma$ through
the classical Einstein relation:
\begin{equation}
D = \frac{2X_0^2}{\beta ^2 \Gamma ^\downarrow \hbar }
  = \frac{T}{\gamma M},
\end{equation}
We note here that our transport theory has a consistent classical limit for
all of these transport coefficients only when they remain finite as
$\hbar \rightarrow 0$. This requires in turn that the parameters of our theory
cannot remain constant as $\hbar \rightarrow 0$, if we are to recover a
well defined classical transport.

\section{ TIME DEPENDENT SOLUTIONS}

In the remaining part of this article we will consider solutions to the
evolution equations (\ref{evol}). For certain forms of the potential,
one can readily obtain explicit solutions to the time evolution of the
density matrix, while for others we solve the evolution equation
numerically.

\subsection{Exactly Solvable Limits}

For the class of potentials:
\begin{equation}
  U(X) = \left\{\begin{array}{cl} -FX & \mbox{Section V}\\
          \frac{1}{2}M\omega^2 X^2& \mbox{Section VI}\\
     -\frac{1}{2}M\Omega^2 X^2 & \mbox{Section VII}\end{array}\right. .
\end{equation}
the evolution equation for the density matrix can be written
as a first order partial differential equation in two variables, which is
readily solved by the method of characteristics\cite{whitham}.

\subsection{Numerical solutions}

For other cases where one does not have analytic results, the method of
solution can be chosen in many ways. For the double well potential, we
have implemented two independent methods which agree. The first is a brute
force technique in which we express the density matrix in the form
\begin{equation}
\rho(X,Y,t) = \sum_{n,m} \alpha_{nm}(t) \phi_m(X)\phi_n^*(Y)
\exp\left [ -\frac{i(E_m-E_n)t}{\hbar}\right ] ,
\end{equation}
where the $\phi(X)$ are the eigenstates in the absence of coupling to
the intrinsic degrees of freedom:
\begin{equation}
  H_0(X)\phi_n(X) = E_n\phi_n(X).
\end{equation}
The evolution equations for $\alpha_{mn}(t)$ are solved numerically in a
truncated $N_{max}^2$ dimensional basis, with $m,n=1,...,N_{max}$.
Typically we will use $N_{max}=31$.

Alternatively, we have solved Eq. (\ref{evol}) on a two dimensional grid
$(X_k,Y_l)$. It is found that the time iteration is most stable using
the alternating--direction implicit method \cite{num}, with five points
interpolation for the space derivatives. As a check on the validity of
the solution, the total probability is verified to be preserved at
unity:
\begin{equation}
\int dX \rho(X,X,t) = 1.
\end{equation}
The latter method allows one to integrate much farther in time.

\section{ LINEAR POTENTIAL AND QUANTUM BROWNIAN MOTION}

The classical picture of a Brownian particle in a constant force field
$F$ and interacting with a heat bath is described by the Langevin
equation for the velocity:
\begin{equation}
 \dot{v} + \gamma v -\frac{F}{M}= f(t)
\end{equation}
Here $\gamma$ is the friction coefficient, and $f(t)$ is Gaussian white
noise. In the long time limit the particle energy equilibrates with
$\langle\!\langle p^2\rangle\!\rangle =TM$ and there is a finite drift
velocity $v_{\infty}=F/\gamma M$. In this section we consider the
dynamics of a quantum particle in a constant force field interacting
with our random matrix bath and contrast it to this classical limit.
The results discussed here should be contrasted with those for the case
$\beta=0$, where we have found that the quantum dynamics is similar to
turbulent diffusion\cite{chaos,bdk_pla}.

\subsection{Exact solution}

Let us consider the case when there is a linear potential acting on the
slow variables
\begin{equation}
H_0(X)=-\frac{\hbar ^2}{2M}\partial _X^2 -FX.
\end{equation}
One can consider the case of a time dependent linear potential as
well, i.e. $U(X,t)=-F(t)X$, with only very slight modifications of the
ensuing formulas. The equation for the density matrix now becomes
\begin{equation}
\left [ i\hbar \partial _t + \frac{\hbar ^2}{M}\partial _r \partial _s
-\frac{i\beta \Gamma ^\downarrow \hbar ^2 }{2X_0M}
G^\prime \left ( \frac{s}{X_0}\right )\partial _s \right ]
\rho (r,s,t) = \{-Fs + i \Gamma ^\downarrow
[ G (\frac{s}{X_0} ) - 1 ]\} \rho (r,s,t)
\label{eveq}
\end{equation}
with the initial condition $\rho (r,s,0) =\rho_0(r,s)$. It is not
necessary to consider a pure state as an initial state, and we allow for
any general initial density matrix. The mixed partial derivative can be
removed by passing to the Fourier transformed equation (see Section
II.D) for $d(s,k,t)$. This satisfies the equation
\begin{equation}
\left \{ \partial _t +
\left [ \frac{k}{M} - \frac{\beta \Gamma ^\downarrow \hbar }{2MX_0}
G^\prime \left ( \frac{s}{X_0}\right )\right ]
\partial _s \right \} d(s,k,t) =
\left \{ \frac{iFs}{\hbar }+ \frac{\Gamma ^\downarrow }{\hbar }
\left [G\left ( \frac{s}{X_0}\right )-1\right ]
\right \} d(s,k,t). \label{deveq}
\end{equation}
Using the method of characteristics for wave equations \cite{whitham}
one can find the solution in parametric form:
\begin{eqnarray}
d(s,k,t) &=& d_0\left ( S(t),k \right )
\exp \left \{ \int _0^t dt^\prime \left [
\frac{iF S(t-t^\prime)}{\hbar } +
\frac{\Gamma ^\downarrow }{\hbar }
\left [G\left (\frac{S(t-t^\prime )}{X_0}\right )-1\right ]
\right ] \right \} ,
\label{dsol} \\
\rho (r,s,t) &=& \int \frac{d k}{2\pi \hbar}
d_0\left (  S(t) ,k \right ) \nonumber \\
& \times & \exp \left \{ \int _0^t dt^\prime \left [
\frac{iF S(t-t^\prime )}{\hbar } +
\frac{\Gamma ^\downarrow }{\hbar }
\left [ G\left ( \frac{S(t-t^\prime)}{X_0} \right ) -1 \right ]
\right ] \right \} ,
\end{eqnarray}
where the time dependent function $S(t-t^\prime)$ is the solution of the
auxiliary equation
\begin{equation}
\frac{d S(\tau )}{d\tau }= -\left [ \frac{k}{M} -
\frac{\beta \Gamma ^\downarrow \hbar }{2MX_0}
G^\prime \left ( \frac{S(\tau ) }{X_0}\right )\right ]. \label{lintraj}
\end{equation}
Here $d(s,k,t)|_{t=0}=d_0(s,k)$ is the initial distribution, which is
just the Fourier transform of $\rho_0(r,s)$. The coordinate $s$ appears
in this solution as an initial condition on $S$: $S(0)=s$.  These
equations define the flow of the density matrix in time. For an
arbitrary $t^\prime$ the function $S(t-t^\prime)$ also satisfies the
homogeneous equation
\begin{equation}
\left \{ \partial _t +
\left [ \frac{k}{M} - \frac{\beta \Gamma ^\downarrow \hbar }{2X_0M}
G^\prime \left ( \frac{s}{X_0}\right )\right ]
\partial _s \right \} S(t-t^\prime) =0.
\end{equation}
Again, $ S(t-t^\prime) $ depends on $s$ through the initial condition.
One can reexpress the full solution in terms of an initial
density matrix $\rho_0$ as well:
\begin{eqnarray}
\rho (r,s,t) &=& \int \!\!\int \frac{dr^\prime d k}{2\pi \hbar }
\rho _0\left ( r^\prime , S(t)  \right ) \nonumber \\
& \times & \exp \left \{ \frac{ik(r-r^\prime)}{\hbar} +
\int _0^t dt^\prime \left [
\frac{iF S(t-t^\prime)}{\hbar } +
\frac{\Gamma ^\downarrow }{\hbar }
\left [ G\left (\frac{S(t-t^\prime)}{X_0}\right ) -1\right ]
\right ] \right \} . \label{dens_lin}
\end{eqnarray}

There are no restrictions on the initial conditions $\rho_0(r,s)$, but it is
convenient in our considerations below to use a particular form. If we have an
initial Gaussian wavefunction,
$\psi_0(X)=\exp{(-X^2/4\sigma^2)}/(2\pi\sigma^2)^{1/4}$, then:
\begin{eqnarray}
\rho_0(r,s) &=&\frac{1}{\sqrt{2\pi\sigma^2}}\exp\left( -\frac{4r^2+s^2}
     {8\sigma^2}\right),\\
d_0(s,k) &=& \exp\left(-\frac{k^2\sigma^2}{2\hbar^2} -
                  \frac{s^2}{8\sigma^2}\right).
\end{eqnarray}

\subsection{Attractors and Repellors}

It is clear that the time evolution of $\rho(r,s,t)$ depends on the
properties of $S(t)$. Hence the flow of the solution $\rho(r,s,t)$ can
be better understood if we examine the stable and unstable fixed points
of $S(t)$. In order to discuss further the character of this solution it
is convenient to use a specific form for the correlator $G$, so for
illustrative purposes we use $G(x)=\exp(-x^2/2)$. The fixed points are
determined from the condition
\begin{equation}
 k = \frac{\beta \Gamma ^\downarrow \hbar }{2X_0}
G^\prime \left ( \frac{s}{X_0}\right )
\end{equation}
which is plotted in Fig. 1 (top). (The analogous result for the case of
a periodic correlation function $G(x)=\cos[x]$ is shown in Fig. 1
(bottom).) This has a maximum value $k_0$ given by
\begin{equation}
k_0= \frac{\beta \Gamma ^\downarrow \hbar }{2\sqrt{e}X_0} \label{k0}.
\end{equation}
The character of the  trajectories $S(\tau)$ determined by solving  Eq.
(\ref{lintraj}) depends on whether $|k|>k_0$ or $|k|\le k_0$. Since $k$
is not dynamical, the evolution is only along the $s$ direction.  The
flow lines $S(\tau)$ are shown in Fig. 1 with the arrows for selected
values of  $k$. As one can see, the part of the curve between $-1< s/X_0
< 1$ is a line of stable fixed points (attractors), while for
$|s/X_0|>1$, it becomes a line of unstable fixed points (repellors).
When  $|k|>k_0$ the r.h.s. of Eq. (\ref{lintraj}) maintains a definite
sign so $S(\tau )$ is either a monotonically increasing or decreasing
function of time, for any given initial condition $S(0)=s$. When
$-k_0\le k \le k_0$ one can see from Fig. 1 that there are two types of
solutions. Furthermore, since $k_0$ depends on $\beta$, as $\beta\rightarrow
0$,  $k_0\rightarrow 0$ and the character of the dynamical evolution
depends on temperature.

In general the trajectory $S(\tau )$ can be determined through a simple
quadratures
\begin{equation}
\tau = \int _s^{S(\tau )}dx \left [ -\frac{k}{M} +
\frac{\beta \Gamma ^\downarrow \hbar }{2MX_0}
G^\prime \left ( \frac{x}{X_0}\right ) \right ]^{-1}.
\end{equation}

\subsection{Momentum Cumulants -- Thermalization}

In order to determine the momentum cumulants we must  construct the
Taylor expansion of the function $\ln d(s,k,t)$ in powers of $s$  at
$k=0$. One can see from Fig. 1 that along the $k=0$ line, all
trajectories flow to the origin $S=0$ in the long time limit. By
changing the integration variable in  Eq. (\ref{dsol}) from time to
$x=s/X_0$,  using Eq. (\ref{lintraj}), in the limit $t\rightarrow \infty
$, where all trajectories have the property that $S(t)\rightarrow 0$  we
obtain
\begin{eqnarray}
\ln d (s,0,t)&=& \ln d_0(S(t),0)+
\frac{1}{i\hbar}\;
\frac{2FMX_0^3}{\beta \Gamma ^\downarrow \hbar }
\int _0^{s/X_0} dx \exp \left ( \frac{x^2}{2} \right ) \label{oscp}\\
&+&  \frac{2MX_0^2}{\hbar ^2 \beta }
\int _0 ^{s/X_0} \frac{dx}{x} \left [ 1 - \exp \left (
\frac{x^2}{2} \right ) \right ] . \nonumber
\end{eqnarray}
From the power series expansion of the integrands, one can readily read
off all the momentum cumulants in the limit $t\rightarrow \infty$. In
this limit the initial condition become irrelevant as $\ln d_0(S(t),0)
\rightarrow 0$. In particular the first and second cumulants are:
\begin{eqnarray}
\langle \!\langle p \rangle \!\rangle &=&
         \frac{2MFX_0^2}{\beta \Gamma^\downarrow \hbar }
         =\frac{F}{\gamma } \\
\langle \!\langle p^2 \rangle \!\rangle &=& \frac{M}{\beta } =MT
\end{eqnarray}

One can see the physical picture emerging here. In the long time limit the
quantum particle reaches a terminal velocity determined from the first
cumulant:
\begin{equation}
v_\infty = \frac{ \langle \!\langle p \rangle \!\rangle }{M}   =
\frac{2FX_0^2}{\beta \Gamma^\downarrow \hbar }=\frac{F}{M\gamma },
\end{equation}
with the definition of the friction coefficient $\gamma$ identical to
that in the Kramers equation, Eq. (\ref{frict}). Furthermore, the kinetic
energy of the particle equilibrates to the proper thermal equilibrium
result:
\begin{equation}
\frac{\langle p^2\rangle }{2M}=
\frac{\langle \!\langle p   \rangle \!\rangle ^2}{2M} +
\frac{\langle \!\langle p^2 \rangle \!\rangle   }{2M}
= \frac{Mv_\infty ^2}{2} +\frac{T}{2}.
\end{equation}

What is more notable, however, is that the momentum distribution has
higher than second order cumulants, which increase exponentially with
the order of the cumulant. In the absence of the linear potential
($F=0$) only the even order cumulants are nonvanishing, and are given
by:
\begin{equation}
\langle \!\langle p^{2n} \rangle \!\rangle = (-1)^{n-1}
\frac{  (2n-1)!!}{n}\frac{MX_0^2}{\hbar ^2\beta}
\left(\frac{\hbar}{X_0}\right)^{2n}.
\end{equation}
The presence of $F$ adds only odd cumulants:
\begin{equation}
\langle \!\langle p^{2n-1} \rangle \!\rangle = (-1)^{n-1} (2n-3)!!
\frac{FX_0}{\gamma \hbar}
 \left(\frac{\hbar}{X_0}\right)^{2n-1}.
\end{equation}
All higher than second order cumulants vanish in the strict classical
limit $\hbar \rightarrow 0$. These cumulants also vanish in the limit
$X_0\rightarrow \infty$, which shall be interpreted as a weak coupling
limit to the thermostat (which is the case in statistical physics). When
the coupling to the ``thermostat'' is not weak, in the $t\rightarrow
\infty $ limit the function $d(s,0,t)$ (which is the Fourier transform
of the momentum distribution) is narrower than a Gaussian and which thus
leads to an equilibrium momentum distribution with longer tails. This is
exemplified in Fig. 2, where we compare the natural logarithm of the
momentum characteristic function Rel. (\ref{oscp}) for the case when
$M=\hbar=\beta =1, \;\; F=0$ with (from narrowest to widest) $X_0=0.1$,
0.5, 1, 2 and $X_0=\infty$ (the Gaussian limit).
The presence of a linear potential does not modify the absolute value of
the characteristic function, only its phase.

Naively, one would have expected that the coupling to the thermostat is
controlled by the magnitude of $\Gamma ^\downarrow$ alone. As one can
easily convince oneself however, the coupling between the two systems is
also controlled by the correlation length $X_0$. In the limit
$X_0\rightarrow \infty$ there is no energy exchange between the two
subsystems, irrespective of the value of the ``coupling constant''
$\Gamma ^\downarrow$. For $X_0=\infty$ the reservoir never responds to
the ``external agent'' and only its excitation spectrum acquires GOE
fluctuation characteristics if $\Gamma^\downarrow$, $\rho _0$ and
$\kappa _0$ satisfy certain well known requirements.

\subsection{Coordinate Cumulants and Diffusion }

For the coordinate distributions, we compute the Fourier transform
$d(s=0,k,t)$.  Since $s$ enters as the initial condition, the solutions
which characterize the spatial information are all trajectories with
initial conditions $S(0)=s=0$ and arbitrary $k$. From Fig. 1, we can see
that there are three regions to consider:  {\it 1}) for $|k| < k_0$ the
trajectories have the property that  $S(\tau\rightarrow \infty )$
approach the attractor exponentially, {\it 2})  for $|k|=k_0$, the
trajectories approach the attractor as an inverse power law and {\it 3})
for $|k| > k_0$, the trajectories diverge linearly in time $S(\tau
\rightarrow \infty )\rightarrow -{\mathrm{sign}}(k)\;\infty $. We shall
analyse next the behavior of the characteristic function $d(s=0,k,t)$ in
these different regimes.

\subsubsection{ $|k|< k_0$}

For small $k$ one can linearize Eq. (\ref{lintraj}) around the origin
and solve the simpler equation:
\begin{equation}
\frac{d S(\tau )}{d\tau }= -\frac{k}{M} -
\frac{\beta \Gamma ^\downarrow \hbar S(\tau )}{2MX_0^2}=-\frac{k}{M} -
\gamma S(\tau )
\end{equation}
whose solution is
\begin{equation}
S(\tau )
=\left( s+\frac{k}{\gamma M}\right)\exp (-\gamma \tau) -\frac{k}{\gamma M}.
\end{equation}
In the $t\rightarrow \infty $ limit, by retaining only terms linear in
time we obtain
\begin{equation}
\ln d (s,k,t)|_{s=0} = \ln d_0(S(t),k) +
 \frac{ik}{\hbar} \;
\frac{2F  X_0^2 t}{\beta \Gamma ^\downarrow \hbar } -
\frac{1}{2} \left(\frac{k}{\hbar } \right )^{2}
\frac{4\Gamma^\downarrow X_0^{2}t}{(\beta\Gamma ^\downarrow )^{2}\hbar}
+ {\cal{O}}(k^3). \label{linf}
\end{equation}
For an initial gaussian wavepacket, $\ln d_0(S(t),k)= -S^2(t)/8\sigma^2 -
k^2\sigma^2/2\hbar^2$. In the long time limit, $S(t)$ approaches the fixed
point $s_1(k)$ of (5.12), where $s_1(k) = -kM/\gamma + o(k^3)$. From this we
can determine  the first two spatial cumulants:
\begin{eqnarray}
\langle \!\langle r \rangle \!\rangle &= &
\frac{2F  X_0^2}{\beta \Gamma ^\downarrow \hbar }\; t =v_\infty t,\\
\langle \!\langle r^2 \rangle \!\rangle &= & \sigma^2 +
\frac{\hbar^2}{4M^2\sigma^2\gamma^2} +
\frac{4X_0^2 }{\beta ^2 \Gamma ^\downarrow \hbar } \; t=r_0^2 + 2Dt.
\end{eqnarray}
The constant term in $\langle \!\langle r^2 \rangle \!\rangle$ has two
contributions. The first term is due to the initial width of the gaussian, 
while the second emerges in the long time limit, and is absent at $t=0$.
Physically we see a consistent picture of the quantum dynamics. The
particle position grows linearly, with the velocity given by the
terminal velocity obtained from $\langle \!\langle p \rangle \!\rangle
$. Furthermore, the average position displays diffusion consistent with
Brownian motion, which can be used to determine the diffusion constant
$D$. This is the same expression for $D$ as obtained from the
fluctuation--dissipation theorem in the classical limit (Kramers
equation) in Section III.

As with the momentum distribution, the coordinate distribution in not
Gaussian, and has longer tails. (In the Brownian motion limit, these
tails vanish; see below.) An analytical explicit construction of the
entire spatial distribution and its cumulant expansion is not quite
trivial, as the characteristic function $d(k,0,t)$ has singularities and
different asymptotic time behavior depending on the value of $k$. In
particular, for small values of $k$ the position of the repellor
$s_2(k)\approx {\mathrm{sign}} (k)X_0 \sqrt{-2\ln |k|}$ is not
analytical around $k=0$. This is an indication that either various
moments of the spatial distribution do not exist (perhaps they are
divergent, as in the case of the Cauchy distribution) or they increase
with time at a rate much faster than linear, similar to the case $\beta
=0$ discussed in Ref. \cite{chaos}. In other words, the function $\ln d
(s,k,t)|_{s=0}$ has some singularities in the $k$--plane, which have to
be dealt with  more carefully.

More generally, for $|k|<k_0$ and large times Eq. (\ref{lintraj}) can be
written approximately as
\begin{equation}
\frac{dS(\tau )}{d\tau }= -\frac{\beta \Gamma ^\downarrow \hbar }{2MX_0^2 }
\exp \left ( -\frac{s_1(k)^2}{2X_0^2} \right )
\left (1 - \frac{s_1(k)^2}{X_0^2}\right ) (S(\tau ) - s_1(k) )  ,
\end{equation}
where $s_1(k)$ is the $k$--dependent position of the attractor and the
boundary condition is in this case $S(\tau \rightarrow \infty
)\rightarrow s_1(k)$. In computing the leading term in $t$ one can use
this approximate trajectory. In this way one arrives at
\begin{equation}
\ln d (s,k,t)|_{s=0} \approx \ln d_0 (s_1(k),k)|_{s=0} +
\left \{ \frac{i F s_1(k) }{\hbar } + \frac{\Gamma ^\downarrow }{\hbar }
\left [ \exp \left ( -\frac{s_1(k)^2}{2X_0^2}\right ) -1 \right ] \right \}
\; t .
\end{equation}
For small values of $k$ the function $d(0,k,t)$ is narrower than a
Gaussian, as one can establish easily by comparing the Taylor series in
$k$ of this expression  with Eq. (\ref{linf}). Since the Fourier
transform of this function is nothing else but the spatial distribution,
we can thus conclude that at large distances the spatial distribution
has longer tails than a Brownian particle.

\subsubsection{ $|k|= k_0$}

For this critical value of $k$, the trajectory with initial condition
$s=0$, $k=\pm k_0$, will approach $S=\mp X_0$ in the limit
$\tau\rightarrow\infty$, as seen in Fig. 1. It is sufficient to look at
the case $k=k_0$. To examine the behavior of $S(\tau)$ in the
neighborhood of the fixed point, it is convenient to take:
\begin{equation}
 S(\tau) = -X_0(1-\varepsilon(\tau)).
\end{equation}
The dynamics is then given by
\begin{equation}
\frac{d\varepsilon(\tau )}{d\tau }= -\frac{k_0}{MX_0} +\gamma
(1-\varepsilon)\exp\left [ -\frac{(1-\varepsilon)^2}{2}\right ] \simeq
-\frac{k_0}{MX_0}\varepsilon^2.
\end{equation}
The solution in terms of $S$ then has the power--law behavior:
\begin{equation}
S(\tau ) = -X_0\frac{k_0\tau}{ k_0\tau + MX_0 }.
\end{equation}
Substituting this into the solution for $d(s=0,k,t)$, we obtain:
\begin{eqnarray}
\ln d (0,k_0,t) &=& \ln d_0(S(t),k_0)+
\left [ \left (\frac{k_0 }{MX_0\gamma }-1\right )\frac{\Gamma
^\downarrow}{\hbar }  - i\frac{FX_0}{\hbar} \right ] t \\
&+&
\left [ \frac{\Gamma ^\downarrow }{\hbar \gamma } +
i\frac{FMX_0^2}{\hbar k_0} \right ] \ln\left[
     1+\frac{k_0}{MX_0}t\right]
           + o(k_0^3). \nonumber
\end{eqnarray}

\subsubsection{ $|k|> k_0$}

In the other regime, when $|k|\gg k_0$ and $t\rightarrow \infty$,
$\beta$ no longer plays a role. Strictly speaking, this analysis is only
valid in the $\beta=0$ limit. The approximate equation for the
trajectory is
\begin{equation}
\frac{dS(\tau )}{d\tau }= -\frac{k}{M},\qquad\qquad S(\tau)=-\frac{k}{M}\tau,
\end{equation}
and thus the characteristic function acquires approximately the form
\begin{equation}
\ln d (s,k,t)|_{s=0} = \ln d_0\left (-\frac{kt}{M},k\right )
-\frac{i F k t^2}{2\hbar M} + \frac{\Gamma ^\downarrow }{\hbar }
\left [ \frac{\sqrt{2\pi} MX_0}{2|k| }  - t \right ]
\;  .
\end{equation}
Notice, that if for $|k|<k_0$ the characteristic function is narrower
than a Gaussian, for $|k|>k_0$ one has just the opposite, the
characteristic function tends to a time--dependent constant ({\it modulo} a
nontrivial phase however).
The above relation shows that the short distance behavior of the propagator
has a quite unusual behavior. In order to construct the propagator one
should use the initial density distribution $\rho _0(r,s) = \delta (r)$
and thus $ d_0(s,k) = 1$). The above behaviour for large $k$ implies
that even at finite times an exponentially small part of the initial
spatial distribution is left at the origin, namely $\exp (-\Gamma
^\downarrow t/\hbar )\delta (r)$.

\subsubsection{ General Behavior for all $k$}

An alternate manner to construct the coordinate distribution $d(0,k,t)$
is by solving numerically Eq. (\ref{lintraj}) for $S(t)$ and
substituting the trajectory directly into Rel. (\ref{dsol}) for
$d(0,k,t)$. In Fig. 3 (top) we compare the numerical solution to Eqs.
(\ref{dsol}) and (\ref{lintraj}) (solid) with the Gaussian limit
obtained by keeping only the second cumulant (dashes). The solutions are
obtained for a  time $t=100$, and $X_0=1$. The results
are all similar. However, if one goes to shorter correlation lengths,
the importance of higher order cumulants is striking. In the bottom
figure we show the same for a much shorter correlation length,
$X_0=0.1$, also at $t=100$. All the remaining parameters are kept at
unity. As one can see, the exact result drops off abruptly at $k=k_0$.
The reason is that for $|k|>k_0$, the function $S(t)$ escapes towards
$-\infty$ linearly in time while the solution for $|k|\leq k_0$ slowly
converges to the stable fixed points. If one factors out the overall
$\exp (-\Gamma ^\downarrow t/\hbar )$ from $|d(s,k,t)|$, the reminder
grows exponentially in time for $|k|\leq k_0$ and tends to a time
independent function for $|k|>k_0$ and as a result the discontinuity at
$|k|=k_0$ becomes thus more pronounced.

\subsection{Brownian Motion Limit}

Even though we have shown that the system equilibrates to the correct
thermal limit, the time evolution towards this equlibrium state is
rather complex. For $k=0$ all trajectories have the same asymptotic
behaviour, $S(\tau \rightarrow \infty )\rightarrow 0$, irrespective of
the initial conditions. It is not difficult however to see that if a
trajectory starts far away from the origin, it will take an
exponentially long time to reach the neighborhood of the origin, as the
r.h.s. of Eq. (\ref{lintraj}) is exponentially small for $|s|\gg X_0$.

One can see from the expressions for the cumulants that if the limits
(see also Ref. \cite{chaos})
\begin{eqnarray}
\frac{\hbar}{X_0} &\rightarrow & 0,\\
\frac{X_0}{\Gamma ^\downarrow} &\rightarrow & 0
\end{eqnarray}
are taken, with the friction coefficient $\gamma $ remaining finite, one
obtains the case of pure classical Brownian motion. All but the first
two cumulants for coordinate and momenta vanish, and one is left with a
Gaussian process. These limits can be achieved also by keeping $\hbar $
finite and thus obtaining the case of a quantum Brownian particle.

\section{ QUADRATIC POTENTIAL }

A classical particle in a harmonic oscillator potential, treated with
the Fokker--Planck or Langevin equations, will thermalize, with the
equilibration given by the virial theorem. When the particle is quantum,
the fluctuations are chaotic and moreover the coupling between the two
subsystems is finite, we observe significant departures  from the this
idealized situation.

\subsection{Exact Solution}

For a particle in a harmonic oscillator potential, we start with
\begin{equation}
H_0(X)=-\frac{\hbar ^2}{2M}\partial _X^2 + \frac{M\omega ^2X^2}{2} .
\end{equation}
The equation for $d(s,k,t)$ can be solved through quadratures, in the
same manner as for the linear potential. We note that the
solution we obtain is not limited to the simple case under
consideration. Analytic solutions are also possible if we want to
include a linear time  independent potential term,  a general quadratic
time dependent potential, and/or  a multi--dimensional treatment.

We shall look for a solution using the same representation of the
density matrix introduced in the previous Section in Rel.
(\ref{dtran}). The equation for density matrix is in this case
\begin{eqnarray}
& & \left [ i\hbar \partial _t +
 \frac{\hbar ^2}{M}\partial _r \partial _s
-\frac{i\beta \Gamma ^\downarrow \hbar ^2 }{2X_0M}
G^\prime \left ( \frac{s}{X_0}\right )\partial _s -M\omega ^2 sr \right ]
\rho (r,s,t) \\\nonumber
&=& \left \{ i \Gamma ^\downarrow
\left [ G \left ( \frac{s}{X_0}\right ) - 1 \right] \right \} \rho (r,s,t)
\label{eveq2}
\end{eqnarray}
and the corresponding equation for the function $d(s,t,k)$ is:
\begin{equation}
\left \{ \partial _t +
\left [ \frac{k}{M} - \frac{\beta \Gamma ^\downarrow \hbar }{2MX_0}
G^\prime \left ( \frac{s}{X_0}\right ) \right ] \partial _s - M\omega ^2 s
\partial _k \right \} d(s,k,t) =
\frac{\Gamma ^\downarrow }{\hbar }
\left [G\left (\frac{s}{X_0}\right ) -1 \right ] d(s,k,t). \label{qev}
\end{equation}
The solution of this equation can be obtained again using the method of
characteristics \cite{whitham}. In this case we will have a
two--parameter solution for $d(s,k,t)$ which will depend on the
functions $S(t)$ and $K(t)$ which satisfy the auxiliary equations:
\begin{eqnarray}
\frac{d S(\tau )}{d\tau }&=& -\left [\frac{K(\tau )}{M} -
\frac{\beta \Gamma ^\downarrow \hbar }{2MX_0}
G^\prime \left ( \frac{S(\tau ) }{X_0}\right ) \right ] ,\\
\frac{d K(\tau )}{d \tau } &=& M\omega ^2 S(\tau )
\end{eqnarray}
with the initial conditions
\begin{equation}
S(\tau =0) =s, \;\;\; {\mathrm{and}}\;\;\;
K(\tau =0 )=k .
\end{equation}
We do not write the full solution here since it is more convenient to
solve the equations in action--angle variables. Before we do so, let us
examine the fixed point structure which will emerge.

\subsection{Attractors and Repellors}

The flow in the ``phase space'' $(S,K)$ is not Hamiltonian in character,
which is not surprising. Some typical trajectories are shown in Fig. 4
for $G(x)=\exp(-x^2/2)$. In this case there is no qualitative difference
between the dynamics of this correlator or a periodic function such as
$G(x)=\cos(x)$. In the top of Fig. 4 we have $\omega=1$, $\gamma=0.1$
while in the bottom the friction is stronger, with $\omega=1$,
$\gamma=0.5$. The general pattern of the trajectories in the
$(S,K)$--plane seems to be  quite simple, in the limit $\tau \rightarrow
\infty$, irrespective of the initial conditions, all trajectories spiral
counterclockwise around the origin (unless the motion is overdamped).
The origin is thus a stable focus. There are no trajectories going away
to infinity in any direction in the ``phase space'' $(S,K)$. The plane
$(S,K)$ is separated into four regions by the two curves
\begin{equation}
S=0 \;\;\; {\mathrm{and}}\;\;\;
\frac{K}{M} -
\frac{\beta \Gamma ^\downarrow \hbar }{2MX_0}
G^\prime \left ( \frac{S }{X_0}\right )=0.
\end{equation}
Each of these lines correspond to ${\dot{K}}(\tau )=0$ and ${\dot{S}}
(\tau )=0$ respectively. Near the focus it is simpler to solve the
linearized equations of motion
\begin{eqnarray}
\frac{d S(\tau )}{d\tau }&=& -\frac{K(\tau )}{M}  -
\frac{\beta \Gamma ^\downarrow \hbar S(\tau )}{2MX_0^2},\\
\frac{d K(\tau )}{d \tau } &=& M\omega ^2 S(\tau ),
\end{eqnarray}
which can be solved analytically. If the condition
\begin{equation}
\omega > \frac{\beta \Gamma ^\downarrow \hbar }{4MX_0^2}
=\frac{\gamma }{2}\label{gamma}
\end{equation}
is fulfilled, then the required trajectories are
\begin{eqnarray}
S(\tau )&=& \left [ s \cos \overline{\omega} \tau -
\frac{2k+M\gamma s}{2M\overline{\omega}}
\sin \overline{\omega} \tau
\right ] \exp \left ( - \frac {\gamma \tau }{2}\right  ) ,\\
K(\tau ) &=&
              \left [
                 k \cos \overline{\omega} \tau +
                   \frac{\gamma k+ 2M\omega ^2s }{ 2 {\overline{\omega}} }
\sin \overline{\omega} \tau
              \right ]
\exp \left ( - \frac {\gamma \tau }{2}\right  )
\end{eqnarray}
where
\begin{equation}
\overline{\omega} = \sqrt{ \omega ^2 - \left (
\frac{\beta \Gamma ^\downarrow \hbar }{4MX_0^2}
   \right ) ^2} =\sqrt{ \omega ^2 -\frac{\gamma ^2}{4} } ,
\end{equation}
The case $\omega < \gamma /2$ is formally similar to the case $\omega
=i\Omega$ to be discussed in the next section. Strictly speaking, in the
first order in $\beta$, in which we have derived our formulas so far,
one has $\overline{\omega}\approx \omega$ and we shall use mostly this
approximation from here on.

\subsection{Solution in Action--Angle Coordinates}

The flows shown in Fig. 4 suggest that action--angle coordinates might be
better suited to the dynamical analysis of this problem. The analytical
construction of the higher order cumulants is somewhat cumbersome and we
shall limit our analysis to some general features. The action--angle
variables are in this case
\begin{eqnarray}
I &=&  \frac{k^2}{2M\omega } + \frac {M\omega s^2}{2},\\
\phi &=& \arctan \frac {k}{M\omega s}.
\end{eqnarray}
Here $I$ is related to the energy. We can now rewrite the evolution
equation Eq. (\ref{qev}) in these variables
\begin{eqnarray}
\left \{ \partial _t -\omega \partial _\phi
+ \gamma \exp\left (- \frac{ I\cos ^2 \phi }{M\omega X_0^2}\right )
\left [ \sin \phi \cos \phi \partial _\phi +
       2I \cos ^2 \phi \partial _I \right ]
\right \} d(I,\phi ,t) & & \nonumber \\
=\frac{\Gamma ^\downarrow }{\hbar }
\left [\exp\left (-\frac{I\cos ^2 \phi }{M\omega X_0^2}\right )
-1 \right ]
d(I,\phi ,t).& &
\end{eqnarray}
The equations for the trajectories acquire the following form
\begin{eqnarray}
\frac{d\Phi (\tau )}{d \tau} &=&
\omega - \frac{\gamma }{2} \sin 2\Phi (\tau )
\exp \left \{ - \frac{ \mathrm{I}(\tau ) }{2 M\omega X_0^2}
[1 +\cos 2 \Phi (\tau ) ]\right \} ,\\
\frac{d \mathrm{I}(\tau )}{d\tau }&=&
-\gamma \mathrm{I}(\tau )[1 +\cos 2 \Phi (\tau )]
\exp \left \{ - \frac{ \mathrm{I}(\tau ) }{2 M\omega X_0^2}
[1 +\cos 2 \Phi (\tau )] \right \}
\end{eqnarray}
with initial conditions
\begin{eqnarray}
\mathrm{I}(0) &=& \frac{k^2}{2M\omega } + \frac {M\omega s^2}{2},\\
\Phi (0) &=& \arctan \frac {k}{M\omega s}.
\end{eqnarray}
Irrespective of the initial conditions the action $\mathrm{I}(\tau )$ is
always a monotonically decreasing function, vanishing in the long time
limit. If the motion is not overdamped (i.e. $\omega >\gamma /2$) the
phase $\Phi (\tau )$ is a monotonically increasing function of time. In
the weak friction limit, when $\gamma \ll \omega$, one can replace the
equations for the trajectories with the following approximate equations
\begin{eqnarray}
\frac{d\Phi (\tau )}{d \tau} &=&
\omega  ,\\
\frac{d \mathrm{I}(\tau )}{d\tau }&=&
-\gamma \mathrm{I}(\tau )
\exp \left [ - \frac{ \mathrm{I}(\tau ) }{2 M\omega X_0^2} \right ]
\left [
{\cal I}_0 \left ( \frac{\mathrm{I}(\tau ) }{2 M\omega X_0^2} \right )-
{\cal I}_1 \left ( \frac{\mathrm{I}(\tau ) }{2 M\omega X_0^2} \right )
\right ] , \label{action}
\end{eqnarray}
obtained after averaging the initial equations over the fast motion
(i.e. over one period $2\pi/\omega $). Here ${{\cal I}}_0 (x)$ and
${{\cal I}}_1 (x)$ are the modified Bessel functions of first kind. The
r.h.s. of  Eq. (\ref{action}) behaves linearly in $\mathrm{I}$ for small
values of the action and as $1/\sqrt{\mathrm{I}}$ for large values. This
behaviour is consistent with the fact that the friction term is
effective only near the origin in coordinate, i.e. for $s\approx
{\cal{O}}(X_0)$.

After manipulations similar to those used in the previous section,
and averaging over the fast motion and by changing the integration
variable from time to action (using Eq. (\ref{action})), one obtains
that the asymptotic expression for the characteristic function is
\begin{eqnarray}
\ln d(I,\phi ,t) &=& \frac{2MX_0^2}{\hbar ^2 \beta }
\int _0^I \frac{d \alpha }{\alpha }\left [
{{\cal I}}_0\left ( \frac{ \alpha }{2M\omega X_0^2} \right ) -
\exp \left ( \frac{ \alpha }{2M\omega X_0^2} \right ) \right ]
\label{osccum} \\
& \times &
\left [ {{\cal I}}_0\left ( \frac{ \alpha }{2M\omega X_0^2} \right ) -
{{\cal I}}_1\left ( \frac{ \alpha }{2M\omega X_0^2} \right ) \right ]
^{-1}
\nonumber
\end{eqnarray}
where $I={\mathrm{I}}(0)$, which was defined above as the initial value
of the action variable. One can construct in a straightforward manner
all the spatial and momentum cumulants, by reverting to the initial
space--momentum variables $s$ and $k$.  Note that $ d(s,k,t)=d(I,\phi
,t)$ is time independent in this limit, as expected. One can extract
easily the behaviour for large and small $I$.
\begin{equation}
\ln d (I,\phi ,t) \approx \left \{
\begin{array}{ll}
-\frac{I}{\hbar^2\omega \beta }
-\frac{3I^2}{16\hbar^2 \beta \omega ^2 MX_0^2}&
   \;\;\;  I \ll M\omega X_0^2 , \\
 & \\
-\frac{8\sqrt{2\pi }MX_0^2}{3\beta \hbar ^2}
\left [\frac{I}{2M\omega X_0^2}\right ]^{3/2} &
 \;\;\; I \gg M\omega X_0^2  .
\end{array}
\right. . \label{oscasym}
\end{equation}
By taking the Fourier transform of the above expression for $d(s,k,t)$
for either $k=0$ or $s=0$ one can determine either the momentum or the
spatial equlibrium distribution of a harmonic oscillator coupled with a
``reservoir''. For the oscillator, the action variable $I$ is, up to a
trivial factor, simply the total energy $E$. Hence one can also extract
the energy distribution from the above expressions.

\subsection{Eigenvalues of the Time Evolution Operator}

It is instructive to construct the eigenvalues and the eigenvectors of
the time evolution operator. If we rewrite the time evolution equation
for the density matrix in the form,
\begin{equation}
 i\hbar\partial_t\rho(r,s,t)={\cal O}\rho(r,s,t)
\end{equation}
we can consider the eigenvalue problem associated with ${\cal O}$:
\begin{equation}
 {\cal O}\rho(r,s)= \lambda \rho(r,s).
\end{equation}
The equilibrium solution will correspond to $\lambda=0$, and in general
the spectrum should be complex. While an analytical solution does not
seem possible for a general $G(x)$, if we take $G(x)=1-x^2/2$, one can
readily solve the problem.

For the equilibrium state $(\lambda=0)$ one obtains
\begin{equation}
\rho_0(r,s) = \exp\left\{ -\frac{\beta M\omega^2r^2}{2} -
\frac{M}{2\beta\hbar^2} s^2\right\}.
\end{equation}
We will discuss the physical properties of this solution in the next
section. This problem now becomes identical to the eigensolutions of the
Fokker--Planck equation for the oscillator\cite{vankampen}. Following that
analysis, the eigenvalue spectrum is given in terms of two integers,
$n_1$ and $n_2$. The basic roots are
\begin{equation}
\lambda_\pm=-\frac{i\gamma\hbar}{2} \pm  \hbar\overline{\omega} ,\qquad\qquad
\overline{\omega} = \sqrt{\omega^2-\frac{\gamma^2}{4}}
\end{equation}
and the entire spectrum can be written as:
\begin{equation}
\lambda_{n_1,n_2} = -\frac{i\gamma\hbar }{2}(n_1+n_2) +
 \hbar\overline{\omega} (n_1-n_2),\qquad\qquad (n_i=0,1,...)
\end{equation}
The time evolution of the density matrix will equilibrate to the
$\lambda=0$ eigenvector, as all other components will decay in time as
$\exp [-(n_1+n_2)\gamma t/2\hbar ]$. The temperature dependence appears
indirectly through $\gamma$, which vanishes in the $\beta=0$ limit.
These patterns are shown in Fig. 5 for $\beta> 0$, where it is clear
that all eigenvalues with the exception of the equilibrium state
$(\lambda=0)$ lead to decay. For $\beta=0$, the spectrum collapses to
the real axis, and all points $\lambda=n\overline{\omega} $ are
infinitely degenerate.

\subsection{Recovery of Equilibrium Thermodynamics}

For the harmonic oscillator, we would like to see whether or not the
random matrix bath can act as an ideal heat bath. The quantum
equilibrium density matrix we would expect is:
\begin{equation}
\rho_{eq}(X,Y) = \sum_{n=0}^\infty
\exp \left [ -\beta \left ( n+\frac{1}{2}\right ) \hbar\omega \right ]
\phi_n(X)\phi_n(Y)
\end{equation}
where $\phi_n$ are the oscillator wavefunctions given by:
\begin{equation}
\phi_n(X) = \left[\frac{\alpha^2}{\pi 2^{2n}(n!)^2}\right]^{1/4}
H_n(\alpha X)\exp \left (-\frac{\alpha^2X^2}{2} \right ),
\qquad\qquad \alpha^2=\frac{M\omega}{\hbar}.
\end{equation}
and $H_n$ is the Hermite polynomial. Defining $z=\exp[-\beta\hbar\omega]$, we
can write:
\begin{eqnarray}
\rho_{eq}(X,Y) &=& \sqrt{\frac{\alpha z}{\pi}}
\exp \left [-\frac{\alpha^2(X^2+Y^2)}{2}\right ]
 \sum_{n=0}^\infty \left(\frac{z}{2}\right)^n\frac{1}{n!} H_n(\alpha X)
 H_n(\alpha Y)\\
 & = & \sqrt{\frac{\alpha z}{\pi (1-z^2)}}\exp\left
\{ -\frac{\alpha^2}{2(1-z^2)}
 \left [ (X^2+Y^2)(1+z^2) -4XYz\right ]\right \} \\\nonumber
\end{eqnarray}
In the last line we have used the generating function for Hermite
polynomials. If we take the leading order in $\beta$ and transform from
$X,Y$ to $r,s$, we find
\begin{equation}
\rho_{eq}(r,s) \simeq\sqrt{\frac{\alpha}{2\pi\beta\hbar}}
 \exp\left[-\frac{\alpha^2}{2\beta\hbar\omega}s^2
 -\frac{\alpha^2\beta\hbar\omega}{2}r^2\right],
\end{equation}
which is precisely the $\lambda=0$ eigenvector found in the previous
section. Hence, to leading order in $\beta$ (and $\hbar$ in a sense, as
$\beta$ enters as $\beta\hbar$), and in the Brownian Motion limit, we
recover the equilibrium density matrix. In general the equilibrium
density matrix will have a non--Gaussian character.

\subsection{Cumulant Expansion}

To get the cumulants, one can use the procedure described
previously (see also Ref. \cite{chaos}) and derive that in the $t\rightarrow
\infty $ limit
\begin{equation}
\frac{\langle \!\langle p^{2} \rangle \!\rangle }{2M} =
\frac{\langle \!\langle M\omega ^2r^{2} \rangle \!\rangle
}{2}=\frac{1}{2\beta}=\frac{T}{2},
\end{equation}
as one might have expected. The higher order cumulants can be obtained
by setting $k=0$ or $s=0$ respectively in the Taylor expansion in $I$ of
the Rel. (\ref{osccum}). We shall not try here to derive  explicit
expressions for higher order cumulants, as they apparently seem to be
quite complicated and hardly very revealing. As Rel. (\ref{oscasym})
clearly shows, the equilibrium distribution is much narrower than a
Gaussian one, which is already indicative of the presence of
nonnegligible higher then second order cumulants.  In Fig. 6 we compare
$\ln d(I,\phi, t\rightarrow \infty )$ as a function of
$\sqrt{I}$ obtained from Eq. (\ref{osccum}) for $X_0=0.5,1,2,\infty$ (narrowest
to widest). The widest $X_0=\infty$ is the equilibrium distribution $\exp
(-\beta H_0)$, (the first term in the low
$I$ expansion, see Rel. (\ref{oscasym})). Notice that the equilibrium
distribution is   independent of $\phi$ as well as of the
actual values of the friction coefficient $\gamma $ and of the coupling
constant $\Gamma ^\downarrow$. Except for a ``trivial'' overall factor,
the shape of this function is controlled by a single parameter, the
``characteristic action'' $2M\omega X_0^2$, which depends on the
``roughness'' of the coupling to the ``reservoir''.

\section{ INVERTED PARABOLIC POTENTIAL }

The tunneling through a potential having the shape of an inverted
parabola can be studied by performing the formal replacement $\omega =
i\Omega _0$ in the corresponding equations of the previous section.

\subsection{Fixed Points}

The equations for the trajectories $(S(\tau ), K(\tau ))$ become now
\begin{eqnarray}
\frac{d S(\tau )}{d\tau }&=& -\left [ \frac{K(\tau )}{M} -
\frac{\beta \Gamma ^\downarrow \hbar }{2MX_0}
G^\prime \left ( \frac{S(\tau ) }{X_0}\right ) \right ] ,\\
\frac{d K(\tau )}{d \tau } &=& -M\Omega _0^2 S(\tau ) . \label{inv}
\end{eqnarray}
The two curves we have introduced in the previous section
\begin{equation}
S=0 \;\;\; {\mathrm{and}}\;\;\;
\frac{K}{M} -
\frac{\beta \Gamma ^\downarrow \hbar }{2MX_0}
G^\prime \left ( \frac{S }{X_0}\right )=0,
\end{equation}
play a similar role in this case as well. However, the general flow
pattern in the ``phase space'' $(S,K)$ has an entirely different aspect,
as shown in Fig. 7. The origin is now an unstable focus point. Near the
origin we can again use the linear approximation $G^\prime (x) \approx
-x$ and thus consider the much simpler equations
\begin{eqnarray}
\frac{d S(\tau )}{d\tau }&=& -  \frac{K(\tau )}{M}  -
\frac{\beta \Gamma ^\downarrow \hbar S(\tau )}{2MX_0^2}  ,\\
\frac{d K(\tau )}{d \tau } &=& - M\Omega _0^2 S(\tau ).
\end{eqnarray}
The imaginary eigenfrequencies are
\begin{equation}
\Omega _\pm = -\frac{\gamma }{2}
\pm \sqrt{ \Omega _0^2 +\frac{\gamma ^2}{4} } =-\frac{\gamma }{2}\pm
\Omega
\end{equation}
with $\Omega _+ \ge 0$ and $\Omega _-<0$. One can easily recognize that
$\Omega _+$ is the renormalized attempt frequency in the classical
Kramers result for the transmission through a barrier \cite{weiss}.

As we have noted in the previous section the case of an inverted
parabolic potential bears some formal similarities with the case of over
damped motion ($\omega <\gamma /2)$, with only one difference, that both
imaginary eigenfrequencies are negative. We shall not discussed this
case in detail in this paper.

For arbitrary initial conditions $S(0)=s,\; K(0)=k$ in the neighborhood
of the origin almost all trajectories  are diverging exponentially away
as $\exp (\Omega _+ \tau )$, except when $s$ and $k$ are exactly along
the line
\begin{equation}
k =  \left ( 1-\frac{\gamma }{2 \Omega }\right ) M\Omega s .
\end{equation}
For sufficiently large $s$ the stable manifold is defined by the line
\begin{equation}
k =  M\Omega _0 s .
\end{equation}
For small friction coefficient ($\gamma \ll \Omega _0$) these two lines
are almost identical. For large values of the friction coefficient
however ($\gamma \gg \Omega _0$) the stable manifold near the origin
almost coincides with the $Os$--axis.

The solution to the above linearized equations of motion are
\begin{eqnarray}
S(\tau )&=&
\left [ \frac{s}{2}\left (1-\frac{\gamma }{2 \Omega }\right )  -
 \frac{k}{2M\Omega }\right ]
\exp \left [ \left ( \Omega -\frac{\gamma }{2}\right ) \tau \right ]
\nonumber \\
& + &\left [ \frac{s}{2}\left (1+\frac{\gamma }{2 \Omega }\right ) +
 \frac{k}{2M\Omega }\right ]
\exp \left [ -\left ( \Omega + \frac{\gamma }{2}\right ) \tau \right ]
,\\
K(\tau ) &=&
\left [ \frac{k}{2}\left  (1+\frac{\gamma }{2 \Omega }\right )
- \frac{M\Omega _0^2 s}{2\Omega } \right ]
\exp \left [  \left ( \Omega -\frac{\gamma }{2}\right ) \tau \right ]
\nonumber \\
& +&\left [ \frac{k}{2}\left  (1-\frac{\gamma }{2 \Omega }\right )
+\frac{M\Omega _0^2 s}{2\Omega } \right ]
\exp \left [ - \left ( \Omega +\frac{\gamma }{2}\right ) \tau \right ] .
\end{eqnarray}
With increasing time and for initial conditions such that $|S(0)=s|\gg
X_0 $, the term responsible for friction in Eqs. (\ref{inv}) becomes
negligible and the trajectories are governed by exponentials $\exp
(\pm \Omega _0\tau )$.

In the case of very small friction ($\gamma \ll \Omega _0$) one can
obtain an approximate expression for the trajectory $S(\tau )$. If one
is interested in the character of the spatial distribution and the way
this is modified in the presence of dissipation, one needs trajectories
with initial conditions $S(0)=0,\; K(0)=k$ only, namely
\begin{eqnarray}
S(\tau )&=& -\frac{k\sinh \Omega _0\tau}{M\Omega _0}
\exp \left [ - \int _0^\tau d \tau ^\prime
 \frac{\tilde{\gamma }(\tau ^\prime ) }{2} \right ],\\
K(\tau )&=& \left [ k\cosh \Omega _0\tau +
\frac{k\tilde{\gamma }(\tau )\sinh \Omega _0\tau }{2\Omega _0}  \right ]
\exp \left [ -\int _0^\tau d \tau ^\prime
\frac{\tilde{\gamma }(\tau ^\prime )}{2} \right ],
\end{eqnarray}
where the ``instantaneous'' friction coefficient is defined as follows
\begin{equation}
{\tilde{\gamma }}(\tau ) = \gamma
\exp \left (
-\frac{k^2\sinh ^2 \Omega _0 \tau }{2M^2\Omega _0^2X_0^2} \right ).
\end{equation}
Using these expressions one obtains for the characteristic function for
the spatial distribution the following expression
\begin{equation}
d(0,k,t) = d_0(S(t), K(t))
\exp  \left  \{   -\frac{\Gamma ^\downarrow }{\hbar}
                \int _0^t d \tau \left  [
\exp \left ( -\frac{S^2(\tau )}{2X_0^2}\right ) -1
                                 \right ]
      \right \}.
\end{equation}
In spite of the fact that it is hopeless to evaluate analytically this
expression as a function of $k$ and time $t$ (there are at least five
exponentials nested into one another), even a cursory analysis of the
$k$--dependence originating from dissipation alone shows that this
function is significantly narrower than a Gaussian. This at once is
indicative of the presence of higher than second order cumulants in the
spatial distribution, but most of all of the fact that the spatial
distribution has a wider spread than the corresponding one obtained in
traditional approaches. If one were to perform a Taylor expansion of
the integrand in powers of $k$ to determine the cumulants for example,
one observes that each power of $k$ is always multiplied by $\sinh
\Omega _0 \tau$. Upon performing the time integral one thus obtains
that all cumulants increase exponentially with time and that in
particular all of them are positive as well.

\subsection{Eigenvalues}

The eigenvalues of the evolution equation are obtained from those of the
harmonic oscillator, by making the transformation $\omega=i\Omega _0$.
Again, specializing to the case $G(x)=1-x^2/2$, we obtain the spectrum:
\begin{eqnarray}
\lambda_\pm & =& -i\hbar\left[\frac{\gamma}{2} \pm
  \sqrt{\Omega^2_0+\frac{\gamma^2}{4}}\right] = i\hbar\Omega_\mp ,\\\nonumber
\lambda_{n_1,n_2} &=& -\frac{i\gamma\hbar}{2}(n_1+n_2) +
i\hbar\sqrt{\Omega^2_0+\frac{\gamma^2}{4}}(n_1-n_2),\qquad\qquad (n_i=1,2,...)
\end{eqnarray}
In this case the eigenvalue $\lambda=0$, which corresponds to an
equilibrium solution, is no longer present. The eigenvalues are all
purely imaginary, and have values above and below the real axis. The
occurrence of eigenvalues with positive imaginary parts is due to the
unphysical asymptotic behaviour of an inverted parabola potential, and
thus to the absence of an equilibrium distribution.

\subsection{Remarks on Tunneling}

Kramers showed that dissipation leads to a reduction of the flux through
an inverted parabolic barrier, as the unperturbed attempt frequency in
the transition state theory should be replaced with the renormalized one
$\Omega _0 \rightarrow \Omega _+\le \Omega _0$ \cite{weiss}. One basic
assumption in Kramers approach was the fact that the friction
coefficient is momentum independent. The present approach can be
interpreted as a theory with a momentum dependent friction coefficient,
which in the zero velocity limit reduces to the classical value. For
finite velocities however, the effective friction coefficient in our
approach is smaller than the one for zero velocity (see the above
approximate solution for the trajectory for the case of small friction).
One can thus expect two effects: {\it i}) the effective or average
attempt frequency in our approach  is in between the Kramers value and
the value corresponding to no friction, i.e. $\Omega _+ < \Omega _{eff}
<\Omega _0$;  {\it ii}) the spatial density distribution is also
modified. Unfortunately the rest of our argument concerning the
behaviour of the spatial distribution is pure qualitative at the present
time. In the classical transition theory the flux through the barrier is
controlled by the probability population of the states at the top of the
barrier. With decreasing friction coefficient, the thermalization of the
spatial distribution in the classically allowed region is less effective
and as a result the overall flux through the barrier decreases as well.
In the previous section we have established that the energy distribution
has generally wider tails than a pure Boltzmann distribution and thus
states with higher energies have a higher relative probability to be
populated when the coupling strength between the ``reservoir'' and the
subsystem is finite. It is thus reasonable to expect that tunneling
states would have a higher relative occupation probability in our
approach. Overall the effect of an effective momentum dependent
friction coefficient is likely to lead to an enhancement of the
tunneling probability when compared with the classical Kramers result.
Since the role of the pure quantum effects and the role of the finite
coupling strength to the ``reservoir'' are intimately related, we cannot
at the present time disentangle the specific role played by each one of
them separately.

\section{ TUNNELING IN A DOUBLE WELL POTENTIAL }

For the problem of tunneling, we consider also the dynamical evolution
of a particle in the double well potential given by:
\begin{equation}
 H_0(X) = -\frac{\hbar ^2}{2M}\partial _X^2 +
 a \left ( X^2-\frac{b}{2a}\right )^2. \label{doublewell}
\end{equation}
For this or more complex potentials, analytic solutions are no longer
possible. The form of the potential is shown in Fig. 8, together with
the first few eigenvalues corresponding to the parameters $a=5$,
$b=M=1/2$.

\subsection{Tunneling Rates}

The numerical solution of the evolution Eq. (\ref{evol}) for the
density matrix is done as explained in Sect. IV.B. Let us examine first
the effect of dissipation on tunneling. For this purpose, we first put
$\beta=0$ and solve the equation for some representative values of the
parameter $ \Gamma^{\downarrow}$. Let $\phi_1(X)$ and $\phi_2(X)$ be the
first two eigenstates of the double well. We take as initial state the
combination
\begin{equation}
 \phi _{0}(X) = \frac{1}{\sqrt{2}} [ \phi_1(X) - \phi_2(X) ],
\end{equation}
which represents a wave packet that is mostly localized in the left well of
Fig. 8. In view of the fact that splitting between the first two states is very
small, $\Delta E \approx 3\times 10^{-6}$, the tunneling of the wave packet to
the other well would take a very long time should there be no dissipation. Fig.
9 shows the strength that tunnels through the barrier as function of the time
for some values of $\Gamma^{\downarrow}$. At the beginning of the tunneling
process, the rate increases almost linearly with this parameter. For very large
time, all the tunneled strengths approach the limiting value 1/2 corresponding
to the equal distribution between the two wells.

\subsection{Temperature Dependence}

Next, we study the dependence of the tunneling on the temperature. For
this, we fix a value of $\Gamma^{\downarrow}$ and solve the evolution
equation for different values of $\beta$. One would expect that the rate
would increase with the temperature, i.e. when $\beta$ decreases. Fig.
10 shows such a behaviour for five values of $\beta = 0., 0.05, 0.10,
0.15, 0.20$. For larger values of $\beta$, the strength on the right
hand side may come out negative for reasons we discussed in Sect. II.

\subsection{Comparison to the Caldeira--Leggett Approach}

At this point, we are also able to compare our results with that
obtained from the Caldeira--Leggett model. To reproduce the latter, it
is sufficient to replace the Gaussian form of the correlation function
that we have used up to now by its quadratic approximation $ G(x)= 1-
x^2/2$ .The effect may be seen better by choosing a smaller value for
the correlation length $X_0$. Fig. 11 shows the density distribution
after the time $t=4$ for the parameters $X_0=0.2,\,\beta=0,\,
\Gamma^{\downarrow}=1 $. The difference is quite appreciable (notice the
ordinate logscale), not only in the tunneled strengths but also in their
shapes.

\subsection{Eigenvalues of the Evolution Operator}

We have numerically computed the eigenvalue spectrum by expanding the
density matrix in the eigenvectors of this quartic potential. Taking
$N_{max}=31$, we have diagonalized the evolution operator ${\cal O}$.
The complex eigenvalues with the smallest imaginary parts are shown in
Fig. 12 (top) for $\beta=0$ and $0.2$. If we denote the eigenvectors as
$\rho_{k}(X,Y)$, then the time evolution, which includes the tunneling,
has the form
\begin{equation}
\rho(X,Y,t) = \sum_k\alpha_k \exp \left ( -\frac{i\lambda_k
t}{\hbar}\right ) \rho_k(X,Y).
\end{equation}
One can see in Fig. 12 that the tunneling can be influenced by several nearby
eigenvalues, and that as $\beta$ increases, the patterns of important
eigenvalues changes considerably. In the bottom of Fig. 12, we plot the
imaginary parts of the eigenvalues  as a function of temperature for the ones
with the smallest imaginary component. One observes that the imaginary part of
the eigenvalue with the  smallest nonvanishing absolute value apparently
vanishes in the limit  $T\rightarrow 0$. This is consistent with Caldeira and
Leggett's conclusion \cite{cald} that at $T=0$ dissipation decreases the
tunnelling rate. This is not true however at high temperatures, as one can
see.

\section{ CONCLUDING REMARKS }

We have developed a dynamical theory of simple quantum systems coupled to
complex quantum  environments, where the environment is a general ``chaotic''
bath of intrinsic excitations. The model Hamiltonian we introduce for the
intrinsic subsystem incorporates the generic properties of finite many--body
systems. This includes an average level density of states sharply increasing
with energy, the presence of universal spectral (or random matrix) fluctuations
for the intrinsic system  and the variation of these properties while changing
the ``shape'' of the intrinsic system modeled with parametric banded random
matrices. In this way, the intrinsic system is capable to easily absorb energy
due to its large heat capacity. We did not yet allow the intrinsic system to
perform mechanical work, but this feature can be readily implemented.

We are able to solve the dynamical evolution equations for these systems,
without making any uncontrollable approximations or assumptions. Using the
Feynman--Vernon path integral approach, we derive the random matrix influence
functional, and an evolution equation for the density matrix of the ``slow''
subsystem at finite, but high temperatures. This evolution equation is
surprisingly easy to manipulate and in many instances one can construct full
solutions by quadratures, for cases when the corresponding path integral can be
computed only by brute force. The entire treatment is performed at the quantum
level. We have shown on the other hand that in the classical limit, the
evolution equation reduces to the Kramers form.

Our analysis was limited to the case when the motion of the simple system can
be treated in the adiabatic approximation. Thus, we have not taken advantage of
another parameter in our description of the reservoir, the bandwidth of the
random matrix $\kappa _0$. It is well known that banded random matrices lead to
localization \cite{izr}. It would be very interesting to explore whether
localization in the ``reservoir'' states induces localization in the simple
quantum system coupled to such a reservoir. (One can show that for finite
bandwidths $\kappa _0$ the influence functional in our model can be brought to
a form similar to the one suggested by Cohen \cite{cohen} to describe
localization, although we did not analyze this limit here.)

The quantum evolution equation we have derived for the density matrix is not
more complicated than a classical Fokker--Planck equation. However, this
evolution equation describes processes which first of all are quantum in nature
and moreover have manifestly non--Gaussian features. So although we  started
with a Gaussian process characterized by our parametric random matrices, we
ended up with a non--Gaussian dynamical process.
Unlike traditional
approaches, we have not assumed that the coupling to the ``reservoir'' is
vanishingly small. In particular, this is one of the reasons why the diffusion
process acquires a non--Gaussian character. The spatial and momentum
distributions are characterized by high order cumulants, which increase
exponentially with the order. As we have discussed earlier \cite{chaos}, this
is indicative of the fact that the corresponding distributions have longer
tails than previously expected. In order to put in evidence such features, one
ought to be careful about the order in which the various limits are taken. If
one takes the semiclassical limit first in the evolution equation for the
density matrix, such features are lost. If in the actual quantum solution one
takes either the semiclassical limit ($\hbar \rightarrow 0$) or the limit of
weak coupling to the reservoir (either $\Gamma ^\downarrow \rightarrow 0$
or/and $X_0\rightarrow \infty$) then one recovers the standard results as well.
We have thus shown that, at least in the framework of the present model (which
we believe to be sufficiently generic), both semiclassical and weak coupling
limits have a rather singular character, which is seen in the temporal domain
or/and the phase space distributions.

The particular form of the quantum evolution equation we have discussed
in this paper has one rather unneccessary restriction, based on the
``translational invariant'' form of the influence functional we have
used, see Eq. (\ref{infl}), which lead to Eq. (\ref{evol}). One can
easily convince oneself that this form we have assumed can be easily
amended and, to within some approximation,
 a more general quantum evolution equation for the density
matrix of the ``slow'' subsystem can be derived, namely:
\begin{eqnarray}
i\hbar \partial _t \rho (X,Y,t) &=&
\left \{ \frac{P_X^2}{2M} - \frac{P_Y^2}{2M} + U(X) - U(Y) \right . \\
 &-& \left .
\frac{\gamma X_0^2 }{2}
\frac{\partial G ( X,Y )}{\partial s }[P_X-P_Y]
+  i \Gamma ^\downarrow
\left [ G (X,Y)-1\right ] \right \} \rho (X,Y,t)
 \nonumber
\end{eqnarray}
where $s=X-Y$ and where the correlator $G(X,Y)$ satisfies the only
symmetry requirements $G(X,Y)=G(Y,X)=G^*(X,Y)$.  One has to interpret
$\gamma$, $\beta$, $\Gamma ^\downarrow$ and $G(X,Y)$ as phenomenological
quantities, characterizing the ``reservoir'' and its coupling to the
``slow'' subsystem. We thus have now at our disposal a quantum evolution
equation, with an effective velocity dependent friction coefficient, a
finite coupling strength to the ``reservoir'' and which in the classical
and weak coupling limits reduces to the Kramers equation. In the weak
coupling limit only this evolution equation is equivalent to the high
temperature limit of the Caldeira--Leggett model.

\acknowledgments
The DOE support for this work is greatly appreciated. The Laboratoire de
Physique Th\'eorique et Hautes Energies is a Laboratoire associ\'e au C.N.R.S.,
URA 0063. Computing facilities provided by IDRIS and NERSC are greatly
appreciated. A.B. thanks H.A. Weidenm\"uller for hospitality and useful
discussions. G.D.D. thanks the Nuclear Theory Group at the University of
Washington for hosting his stay in Seattle. D.K. acknowledges the hospitality
extended to him by W. Haxton at  the Institute for Nuclear Theory and
especially the organizers of the Spring 1997 program, O. Bohigas, A.J. Leggett
and S. Tomsovic, for supporting one of his stays at the INT.

\newpage

\appendix

\section{ Evolution equations for the average propagator and the
influence functional}

The time evolution of the fast subsystem is found by solving the
time--dependent Schr\"{o}dinger equation in the form
\cite{bdk_ann,bdk_mb8,bdk_pre,chaos,bdk_pla}:
\begin{equation}
\phi (t ) = {\mathrm{T}}
\exp \left [-\frac{i}{\hbar } \int _0 ^{t}\! d s H_1(X(s))
 \right ]\phi (0)= {\cal {U}}(X(t))\phi (0). \label{eq:psi}
\end{equation}
where $\mathrm{T}$ is the time ordering operator, and ${\cal {U}}(X(t))$
the propagator. (We assume that the initial state $\phi(0)$ is
uncorrelated with the Hamiltonian $H_1(X(t))$ at later times; correlated
initial conditions have been discussed elsewhere \cite{bdk_ann}.) One
can show that in the leading order in an expansion in $1/N_0$ the
average propagator $U(X(t))=\overline{{\cal U}(X(t))}$ is diagonal in
the representation we have chosen. Its diagonal matrix elements have the
following form
\begin{equation}
U_k(X(t)) = \overline{ \langle k |
\mathrm{T}\exp \left [ -\frac{i}{\hbar}\int _0^t ds
H_1(X(s)) \right ] | k \rangle } =
\exp \left ( -\frac{i\varepsilon _k t}{\hbar } \right )
\sigma (X(t))
\end{equation}
(note that $\sigma (X(t))$ is state independent) and $\sigma (X(t))$
satisfies the following integral equation \cite{bdk_pre}:
\begin{equation}
\sigma(t,t_0) = 1 - \frac{\Gamma ^\downarrow }{\hbar }
 \int _{t_0}^{t} \! \! \! d s_1 \! \int _{t_0}^{s_1} \! \! \! d s_2
\sigma(s_1,s_2)\sigma(s_2,t_0)
P_\beta (s_1 -s_2)
 G \left ( \frac{X(s_1) -X(s_2)}{X_0}\right ) . \label{propag}
\end{equation}
In the following we shall consider that $t_0=0$. Here $P_\beta (s)$ is
the fourier transform of the matrix  band form-factor in the
correlator $\overline{ [h_1(X)]_{ij}[h_1(Y)]_{kl} }$, defined
in Eq. (\ref{correl}):
\begin{equation}
P_\beta (s) = P_\beta ^*(-s) = \frac{\kappa _0}{\sqrt{2\pi } \hbar }
\exp \left [ -\frac{\kappa _0 ^2}{2\hbar ^2}
\left ( s + i\frac{\hbar \beta }{2} \right ) ^2 \right ] , \label{pes}
\end{equation}
 The influence functional can be determined by
solving the following evolution equation:
\begin{eqnarray}
{\cal {L}}(X(t_1),Y(t_2) )&=& \sigma (t_1,0)\sigma ^*(t_2,0)+
\frac{\Gamma ^\downarrow}{\hbar }
\int _0^{t_1}\! d s_1\! \int _0^{t_2}\! d s_2
{\cal {L}}(X(s_1),X(s_2) ) \label{genocc} \\
& &\times P_\beta ^* (s_1 - s_2 )G
\left ( \frac{X(s_1) -Y(s_2) }{X_0}\right )
\sigma (t_1,s_1)\sigma ^*(t_2,s_2). \nonumber
\end{eqnarray}

\section{ Average Propagator at Finite Temperatures }

For a given path $X(t)$ the averaged propagator $\sigma $ satisfies the
following equation
\begin{equation}
\sigma (t,t_0) = 1 -
\frac{\Gamma ^\downarrow }{\hbar } \int _{t_0}^{s_1}\! \! \! d s_1 \!
\int _{t_0}^{s_1} \! \! \! d s_2
\sigma(s_1,s_2)\sigma(s_2,t_0)
P_\beta (s_1 -s_2)
 G \left ( \frac{X(s_1) -X(s_2) }{X_0}\right ) ,
\end{equation}
where
\begin{equation}
P_\beta (s)= \frac{\kappa _0}{\sqrt{2\pi } \hbar }
\exp \left [ -\frac{\kappa _0 ^2}{2\hbar ^2}
\left ( s + i\frac{\hbar \beta }{2} \right ) ^2 \right ] =
P_0(s) +\frac{i\beta \hbar }{2} \frac{d P_0(s)}{ds}+{\cal{O}}(\beta ^2).
\end{equation}
In the above equations $\sigma(s_1,s_2)$ represents the averaged
propagator from time $s_2$ to time $s_1$, as for an arbitrary path
$X(t)$ it is not obvious that this propagator depends only on the time
difference $s_1-s_2$, as was the case in the adiabatic limit and for
$\beta =0$. We shall assume that the following replacement
\begin{equation}
P_0(s) \equiv
\frac{\kappa _0}{\sqrt{2\pi}\hbar }
\exp\left[-\frac{\kappa _0^2 s^2}{2\hbar ^2}\right]
\rightarrow \delta(s) \label{approx}
\end{equation}
is legitimate and compute the first correction in $\beta$ to the
propagator.

In the equations for the propagator and the influence functional
$P_\beta (s)$ enters under integrals with some arbitrary functions
as follows
\begin{eqnarray}
\int _{-\infty} ^0ds P_\beta (s) F(s) &\approx &
\int _{-\infty} ^0 ds P_0 (s) F(s) +
\frac{i\beta \hbar }{2}
\int _{-\infty} ^0ds  \frac{d P_0(s)}{ds}F(s)  \nonumber \\
 &\approx & \frac{1}{2} F(0) +
\frac{i\beta \kappa _0}{2\sqrt{2\pi}}F(0)
-\frac{i\beta \hbar }{4}\left . \frac{dF(s)}{ds} \right | _{s=0} ,
\end{eqnarray}
where we have used the following relations
\begin{eqnarray}
\int _{-\infty} ^{0} ds P_0(s)F(s) &=&\frac{1}{2}F(0) ,\\
\int _{-\infty }^{0} ds \frac{dP_0(s)}{ds} F(s)&=&
\frac{\kappa _0}{\sqrt{2\pi} \hbar }F(0) -
\left . \frac{1}{2}\frac{dF(s)}{ds}\right | _{s=0}, \\
P_0(0)&=&\frac{\kappa _0}{\sqrt{2\pi}\hbar }.
\end{eqnarray}
After a few simple manipulations and after taking into account that
\begin{eqnarray}
\sigma (s,s)&=&1,\\
\left . \frac{d\sigma (s_1,s_2)}{ds_1}\right | _{s_1=s_2}&=&
 \left . \frac{d\sigma (s_1,s_2)}{ds_2}\right | _{s_1=s_2}=0, \\
\left . \frac{ d}{ds_2}G\left (\frac{X(s_1)-X(s_2)}{X_0}\right )
\right | _{s_1=s_2} &=& \left .
\frac{ d}{ds_1}G\left ( \frac{X(s_1)-X(s_2)}{X_0} \right )
\right | _{s_1=s_2}=0,
\end{eqnarray}
one can easily show that up to terms of order ${\cal{O}}(\beta ^2)$  the
averaged propagator is given by the following expression
\begin{equation}
\sigma (t,0)=\sigma(t)=
\exp \left [ -\frac{\Gamma ^\downarrow t }{2\hbar} \left (
 1 +\frac{i\beta\kappa _0}{\sqrt{2\pi}}
-\frac{i\beta \Gamma ^\downarrow}{4} \right ) \right ] .
\end{equation}

In the equation for the influence functional we shall need the quantity
\begin{eqnarray}
\sigma (t) \sigma ^* (t^\prime ) &=&
\exp \left [ -\frac{\Gamma ^\downarrow (t+t^\prime )}{2\hbar}
 +\left ( \frac{i\beta
\kappa _0}{\sqrt{2\pi}} -\frac{i\beta \Gamma ^\downarrow}{4} \right )
\frac{\Gamma ^\downarrow (t-t^\prime) }{2\hbar} \right ] \nonumber \\
&\approx & \exp \left [ -\frac{\Gamma ^\downarrow (t+t^\prime )
}{2\hbar}\right ].
\end{eqnarray}
We are allowed to neglect the imaginary contribution in the exponent as
in all relevant integrals $t-t^\prime ={\cal{O}}(\beta )$.

\section{ Influence Functional at Finite Temperatures }

In order to simplify somewhat the derivation we shall introduce the
``reduced'' influence functional at temperature $\beta $
\begin{equation}
\Lambda _\beta (X(t_1),Y(t_2)) = {\cal{L}}_\beta (X(t_1),Y(t_2))
\exp \left [ \frac{\Gamma ^\downarrow (t_1+t_2)}{2\hbar}\right ] ,
\end{equation}
which, up to correction terms of order ${\cal{O}}(\beta^2 )$ in the
averaged propagators, satisfies the equation
\begin{eqnarray}
\Lambda _\beta (X(t_1),Y(t_2))& = &1 \\
&+&
\frac{\Gamma ^\downarrow }{\hbar}
\int _0^{t_1}\! d s_1\! \int _0^{t_2}\!d s_2
\Lambda _\beta (X(s_1),Y(s_2))P_\beta ^* (s_1 - s_2)G \left ( \frac{X(s_1)
-Y(s_2)}{X_0}\right ). \nonumber
\end{eqnarray}
We have shown earlier \cite{bdk_pre} that for $\beta=0$
\begin{equation}
\Lambda _0 (X(t_1),Y(t_2)) = \Lambda _0 (t_1,t_2)=
\exp \left \{ \frac{\Gamma ^\downarrow }{\hbar }
\int _0^ {t_<} \left
[ G \left ( \frac{X(s)-Y(s)}{X_0} \right ) \right ] ds \right \} ,
\end{equation}
where $t_{<}=\min (t_1,t_2)$. In computing the first order corrections
in $\beta$ for $\Lambda _\beta (X(t_1),Y(t_2))$, we shall proceed as in
the previous appendix, by making an expansion
\begin{equation}
P_\beta ^* (s) \approx P_0(s)-\frac{i\beta\hbar }{2} \frac{dP_0(s)}{ds}
\end{equation}
and taking the limit $\kappa _0 \rightarrow \infty$, which allows us to
make the replacement $P_0(s)\rightarrow \delta(s)$ in all the integrals.
The only term which requires a more careful treatment in the equation
for the reduced influence functional is
\begin{equation}
\beta \frac {\partial \Lambda _\beta (X(s_1),Y(s_2))}{\partial
(s_1-s_2)} \approx \beta \frac {\partial \Lambda _0
(X(s_1),Y(s_2))}{\partial (s_1-s_2)}
= {\cal{O}}(\beta^2 ),
\end{equation}
which can thus be neglected. The reason is that $\Lambda _0
(X(s_1),Y(s_2))$ has a discontinuous partial derivative at $s_1-s_2=0$.
Remembering that we need the influence functional for  $t_1=t_2=t$ only
and by using the obvious exact representation of the integral term
\begin{equation}
\int _0^{t}\! d s_1\! \int _0^{t}\! d s_2 F(s_1,s_2) =
\int _0^{t}\! d s_1\! \int _0^{s_1}\! d s_2 F(s_1,s_2)
+\int _0^{t}\! d s_2\! \int _0^{s_2}\! d s_1 F(s_1,s_2)
\end{equation}
for an arbitrary function $F(s_1,s_2)$ and by applying the rules
described in the previous appendix we obtain that
\begin{eqnarray}
\Lambda _\beta (X(t),Y(t)) &=& 1 +
\frac{\Gamma ^\downarrow }{\hbar}
\int _0^t\! d s \Lambda _\beta (X(s),Y(s))
G \left ( \frac{X(s)-Y(s)}{X_0}\right ) \nonumber \\
& + & \frac{i\beta \Gamma ^\downarrow  }{4X_0}
\int _0^{t}\! d s \Lambda _\beta (X(s),Y(s))
 [ \dot{X}(s)+\dot{Y}(s)]
G^\prime \left ( \frac{X(s)-Y(s)}{X_0}\right )  ,
\end{eqnarray}
where $G^\prime (x) = dG(x)/dx$.
In the above evolution equation one can use either
$\Lambda _0 (s_1,s_2)$ or $\Lambda _\beta (X(s),Y(s))$ on the r.h.s.
An alternative way to derive this equation is to use the fact
that for two given paths $X(s_1)$ and $Y(s_2)$ one has the explicit
symmetry $\Lambda _0 (s_1,s_2)=\Lambda _0 (s_2,s_1)$ and also that
\begin{equation}
\frac{dP_0(s)}{ds}=-\frac{dP_0(-s)}{ds}
\end{equation}
before making the replacement $P_0(s)\rightarrow \delta(s)$ as
described earlier.

This evolution equation can be easily solved as in the case of $\beta=0$
and the final answer for the influence functional is
\begin{eqnarray}
{\cal{L}}(X(t),Y(t)) &=&
\exp \left \{
\frac{\Gamma ^\downarrow }{\hbar } \int _0^ t ds \left [ G\left (
\frac{X(s)-Y(s)}{X_0} \right )- 1 \right ] \right \} \nonumber \\
&\times & \exp \left \{   \frac{i\beta \Gamma ^\downarrow }{4X_0}
\int _0^ t ds [
\dot{X}(s)+\dot{Y}(s)] G^\prime \left ( \frac{X(s)-Y(s)}{X_0}\right )
\right \} .
\end{eqnarray}

\section{ Effective Hamiltonian at Finite Temperatures }

Having obtained an expression for the influence functional at finite
temperatures we can write down from the Feynman--Vernon path integral
an expression for the effective Lagrangian:
\begin{eqnarray}
{\bbox{\mathrm{L}}} &=&
\frac{M\dot{X}^2}{2} - U(X) -\frac{M\dot{Y}^2}{2} + U(Y)\nonumber \\
& &
- i\Gamma ^\downarrow \left [ G\left ( \frac{X-Y}{X_0}\right ) -1\right ]
+\frac{\beta \Gamma ^\downarrow \hbar}{4X_0}[\dot{X}+\dot{Y}]
G^\prime \left ( \frac{X-Y}{X_0}\right ) .
\end{eqnarray}
We can introduce now the corresponding canonical conjugate momenta
\begin{eqnarray}
P_X & = & \frac{\partial {\bbox{\mathrm{L}}} }{\partial \dot{X}} =
M \dot{X} +\frac{\beta \Gamma ^\downarrow \hbar}{4X_0}G^\prime \left (
\frac{X-Y}{X_0}\right ) ,\\
P_Y & = & \frac{\partial {\bbox{\mathrm{L}}}  }{\partial \dot{Y}} =
-M \dot{Y} +\frac{\beta \Gamma ^\downarrow \hbar}{4X_0}G^\prime \left (
\frac{X-Y}{X_0}\right ) \
\end{eqnarray}
and construct the effective Hamiltonian according to usual rules
\begin{eqnarray}
{\bbox{\mathrm{H}}}  &=& P_X \dot{X} + P_Y \dot{Y} -
{\bbox{\mathrm{L}}}  \nonumber \\
 & = & \frac{1}{2M}\left [P_X - \frac{\beta \Gamma ^\downarrow
 \hbar}{4X_0}G^\prime \left (
\frac{X-Y}{X_0}\right ) \right ]^2 + U(X)
 - \frac{1}{2M}\left [P_Y - \frac{\beta \Gamma ^\downarrow
 \hbar}{4X_0}G^\prime \left (
\frac{X-Y}{X_0}\right ) \right ]^2 - U(Y) \nonumber \\
 & + & i\Gamma ^\downarrow  \left [ G\left (
\frac{X-Y}{X_0}\right ) -1 \right ] .
\end{eqnarray}
The requantization  of this effective Hamiltonian is straightforward
($P_X\rightarrow -i\hbar \partial _X$ and $P_Y\rightarrow -i\hbar
\partial _Y$) and it is convenient to reorder the different terms in the
Hamiltonian as  follows:
\begin{eqnarray}
{\bbox{\mathrm{H}}}  &=&
\frac{(P_X+P_Y)(P_X-P_Y)}{2M} + U(X)-U(Y) \nonumber \\
 &-& \frac{\beta \Gamma ^\downarrow \hbar}{4MX_0}G^\prime \left (
\frac{X-Y}{X_0}\right ) (P_X-P_Y)
 +i\Gamma ^\downarrow  \left [ G\left (
\frac{X-Y}{X_0}\right ) - 1 \right ] .
\end{eqnarray}
Any other choice of ordering leads to an evolution equation for the
density matrix $\rho (X,Y,t)$ which does not conserve probability. One
can consider alternative orderings, but in the final analysis these lead
to slight renormalizations of various quantities, but to no qualitative
effects.


\newpage

\noindent {\bf Table I.}  The definition of our parameters as well as some of
the limits used in this analysis.

\vspace{1cm}


\noindent \underline{Parameters:}

\begin{enumerate}
 \item[$X_0$]  Characteristic scale over which $H_1(X)$ changes.
 \item[$N$]    Dimension of the Hilbert space of the intrinsic subsystem. We
               take the limit $N\gg 1$.
 \item[$\kappa_0$] The bandwidth of the random matrix $H_1(X)$.
                The average number
                of states coupled together at a given excitation energy is
                determined by the density of states: $N_0\sim
                \kappa_0\rho(\varepsilon)$. For $\kappa_0\rightarrow\infty$, we
                recover the full random matrix limit.
 \item[$\Gamma^\downarrow$ ] Spreading width. For an initially uncorrelated
                 state evolving under a random matrix, the average
                 propagator decays as $\sigma (t)\sim
                 \exp (-\Gamma^\downarrow t/2\hbar )$.
 \item[$\beta=1/T$]   Inverse thermodynamic temperature, defined through the
                density of states.
 \item[$\rho(\varepsilon)$] The density of states. We use the form
   $\rho(\varepsilon)=\rho_0\exp (\beta \varepsilon )$. When $\beta=0$, the
   intrinsic system has a constant level density.
 \item[$G(x)$]  Correlation function for the intrinsic states. It describes how
  far one must go in $X$ before the intrinsic states are statistically
  uncorrelated. Typically one can use $G(x)=\exp (-x^2/2)$, $1-x^2/2$
  or $\cos x$.
 \item[$\gamma$] Friction coefficient obtained in the classical limit and in
              the full quantum dynamical solution, given by
            $\gamma=\beta\Gamma^\downarrow\hbar/2MX_0^2$.

 \item[$D$] Diffusion constant, given by
              $D=2X_0^2/\beta^2\Gamma^\downarrow\hbar$.
\end{enumerate}


\noindent \underline{Limits:}

\begin{enumerate}
\item[ $ X_0\rightarrow \infty$] Weak coupling limit.
\item[ $ \left.\begin{array}{l}\hbar/X_0\rightarrow 0\\
             X_0/\Gamma^\downarrow\rightarrow 0\\
          \gamma = {\mathrm{finite}}\end{array}\right\}$ ]
            Brownian motion limit.
\item[ $ \hbar \rightarrow 0$] Classical Limit.
\item[ $\kappa_0 \rightarrow \infty$] Adiabatic Limit.
This implies that the collective time scale $X_0/V$ is much longer
than the intrinsic one given by $\hbar/\kappa_0$.
\end{enumerate}

\newpage
\begin{center}
{\bf Figure Captions}
\end{center}

\begin{enumerate}
\item[Figure 1.] Time dependent flow $S(t)$ associated with the time
evolution of the density matrix in a linear potential. (top) The  case
of a Gaussian correlation $G(x)=\exp[-x^2/2]$. (bottom) The flow for a
periodic correlation function $G(x)=\cos[x]$. The stability of the lines
of fixed points are indicated by the directions of the arrows.

\item[Figure 2.] Fourier transform of the momentum distribution,
$d(s,0,t)$, as a function of $s$, for $F=0$ and $M=\hbar=\beta=1$. From
narrowest to widest we have $X_0=0.1,0.5,1,2,\infty$. The $X_0=\infty$
(widest curve) is a Gaussian distribution.

\item[Figure 3.] Fourier transform of the coordinate distribution
$d(0,k,t)$ for the case of a linear potential. The solid curve is the
result obtained from Eqs. \ref{dsol} and \ref{lintraj} from integration
to $t=100$. This is compared to a Gaussian (dashes) obtained from the
second cumulant alone. The top and bottom correspond to different
correlation lengths $X_0=1$ and $X_0=1/10$.

\item[Figure 4.] Time dependent flow $(S(t),K(t))$ associated with the
time evolution of the density matrix in a harmonic oscillator potential,
with the correlation function $G(x) = \exp[-x^2/2]$. The values
of the parameters are $\omega=1$ and $\gamma=0.1$ (top) and $\gamma=0.5$
(bottom).

\item[Figure 5.] Eigenvalue spectrum for the time-evolution operator of
the density matrix $\rho(r,s,t)$ for the harmonic oscillator with
$\beta>0$. For $\beta=0$, the friction $\gamma$ vanishes, and all
eigenvalues lie on the real axis. The $\lambda=0$ solution corresponds
to the equilibrium density matrix.

\item[Figure 6.] Fourier transform of the momentum and coordinate
distribution (they coincide in this case) for the harmonic oscillator
for the case $M=\hbar=\omega=\beta=1$ here represented as a function of
$\sqrt{I}$, where $I$ is the action.  The curves correspond to
$X_0=0.5,1,2,\infty$ from the narrowest to the widest. The Gaussian
distribution $(X_0=\infty)$, obtained from the Brownian motion limit in
which only the quadratic cumulant is non--zero, is shown as a dashed
curve.

\item[Figure 7.] Time dependent flow associated with the time evolution
of the density matrix in an inverted harmonic oscillator potential, with
the correlation function $G(x)= \exp(-x^2/2)$. The parameters are
$\omega=1/10$ and $\gamma=1/2$. The flow of certain trajectories are
shown with the arrows.

\item[Figure 8.] Double well potential and eigenstates in the absence of
the coupling to the intrinsic degrees of freedom.

\item[Figure 9.]  Strengths tunneled to the right hand well for
$\Gamma^{ \downarrow}$ = 0.5, 1 and $\pi $ (upward from the lowest
curve) and $\beta = 0$.

\item[Figure 10.] Strengths tunneled to the right hand well for
$\Gamma^{\downarrow} = \pi $ and $\beta = 0, 0.05, 0.10, 0.15, 0.20 $
(downward from the highest curve).

\item[Figure 11.] Density distributions at t=0 (the narrowest curve) and
at time t=4 for $X_0=0.2,\, \beta=0,\, \Gamma^{\downarrow}=1$: the
continuous (dotted) curve corresponds to using the quadratic (Gaussian)
correlation function.

\item[Figure 12.] (Top) Temperature behavior of the complex eigenvalues
of the double well potential with the smallest imaginary parts.
(Bottom) Temperature dependence of the imaginary part of the eigenvalues
with smallest absolute value. One of the eigenvalues apparently vanishes
in the limit  $T\rightarrow 0$, which suggests that at $T=0$ dissipation
decreases the tunnelling rate\cite{cald}. One can see that the tunneling
process is not always dominated by a single eigenvalue.

\end{enumerate}

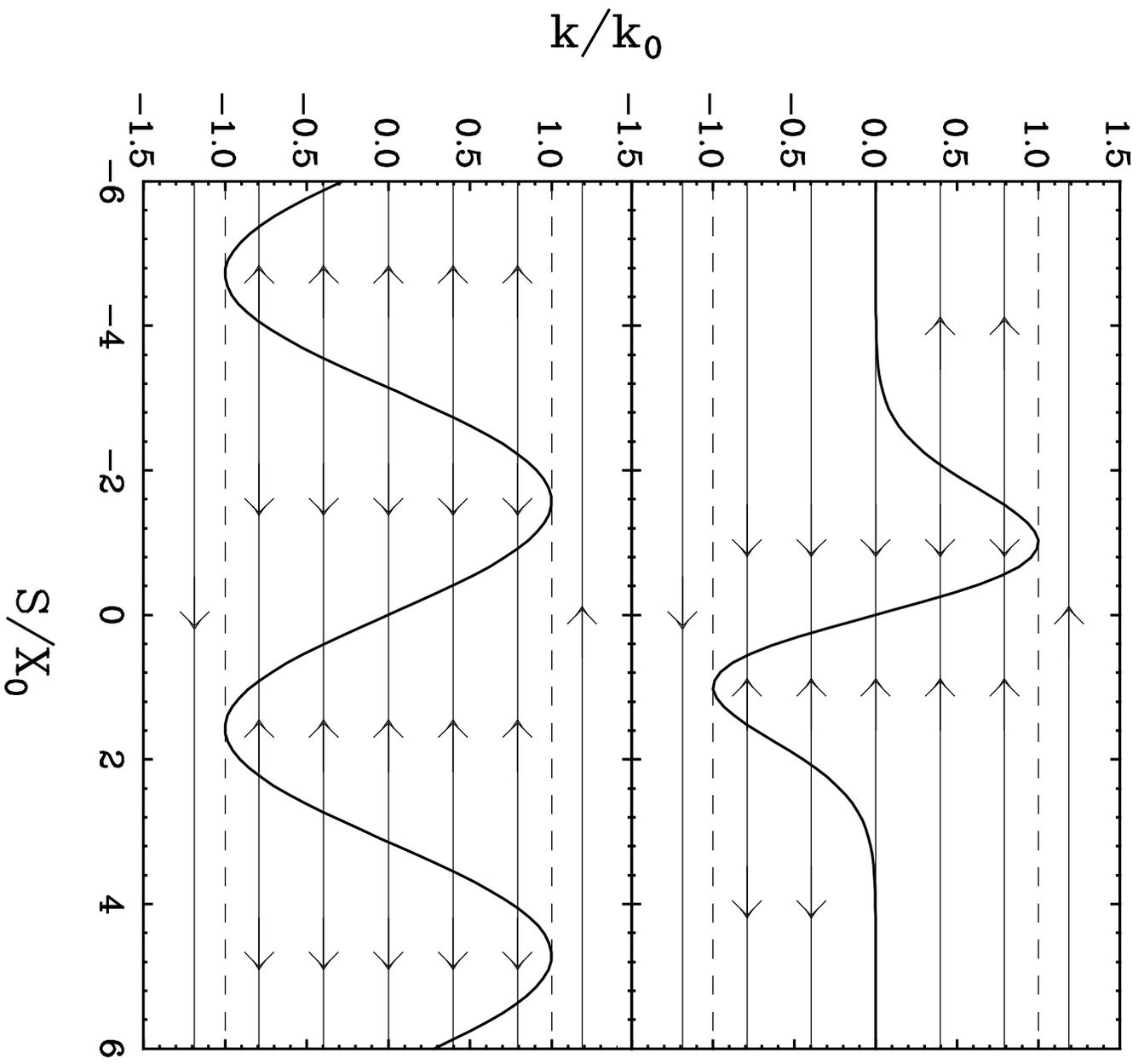
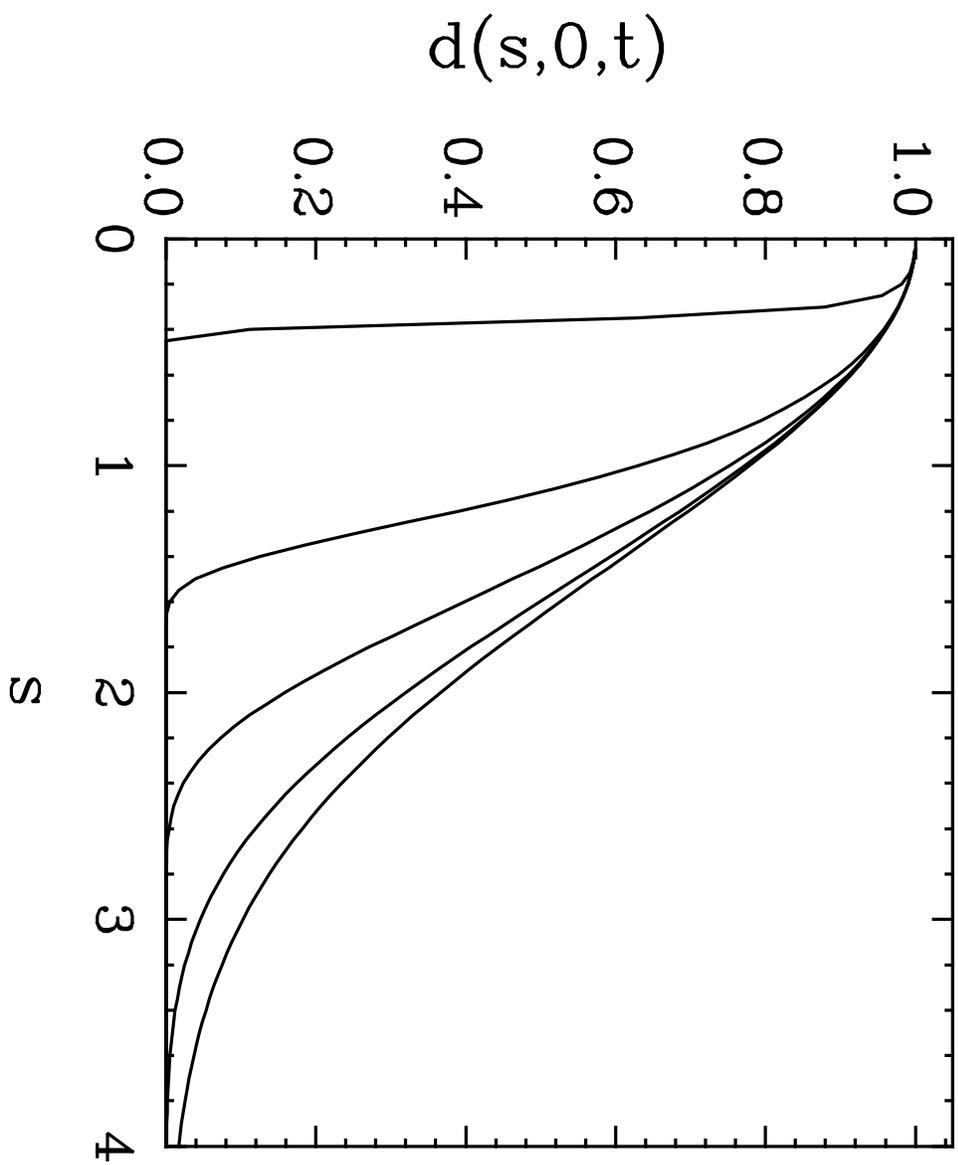
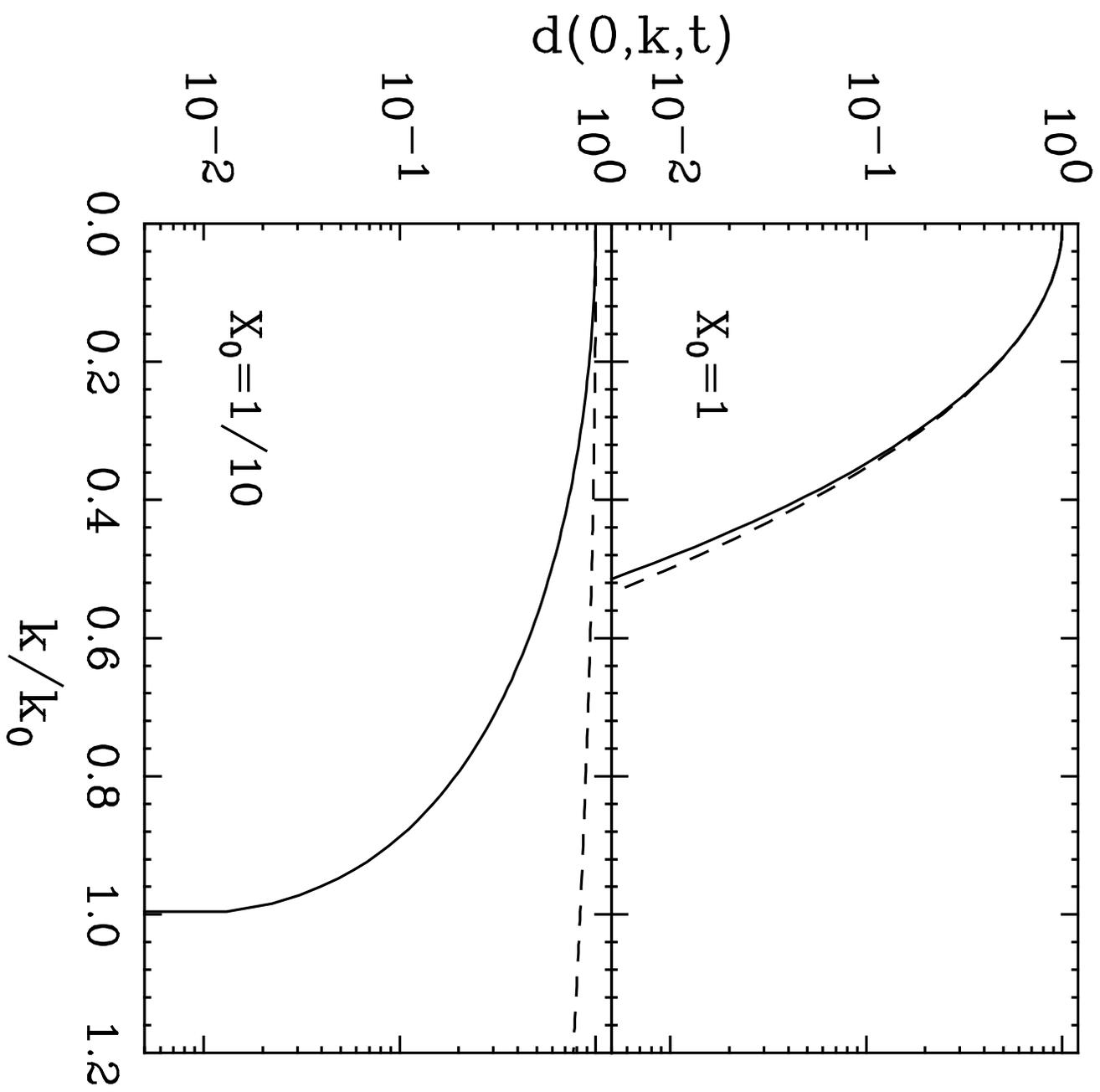
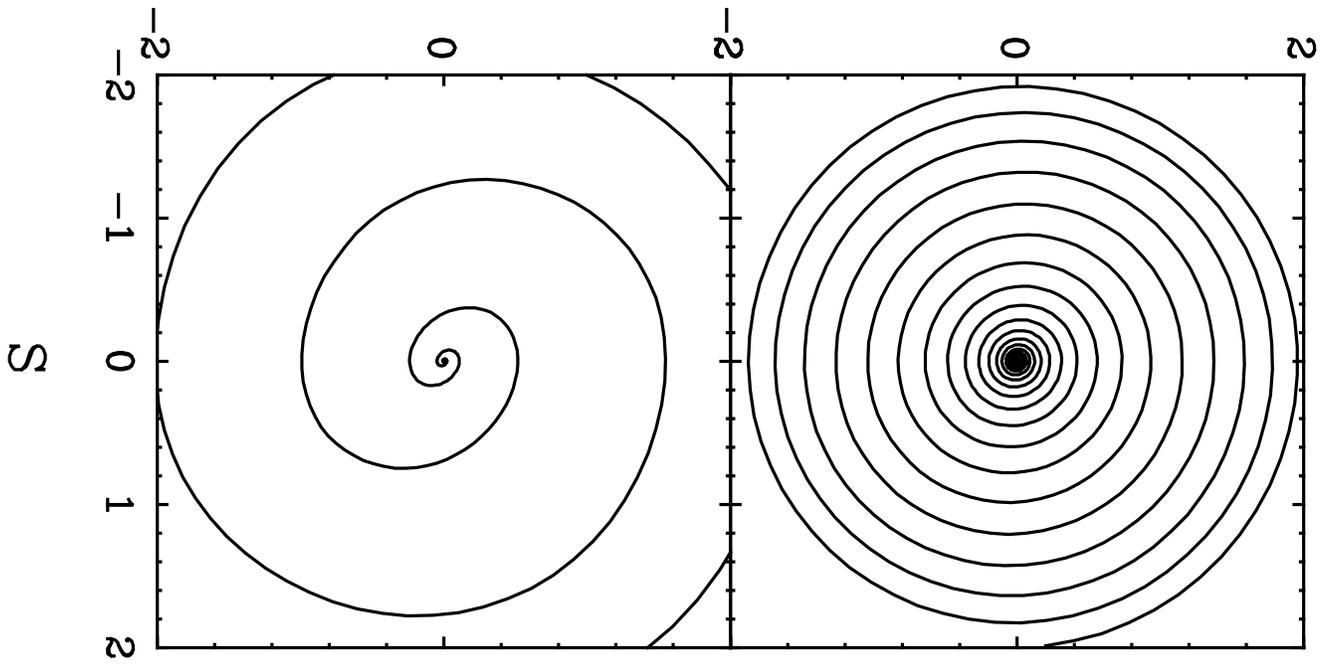
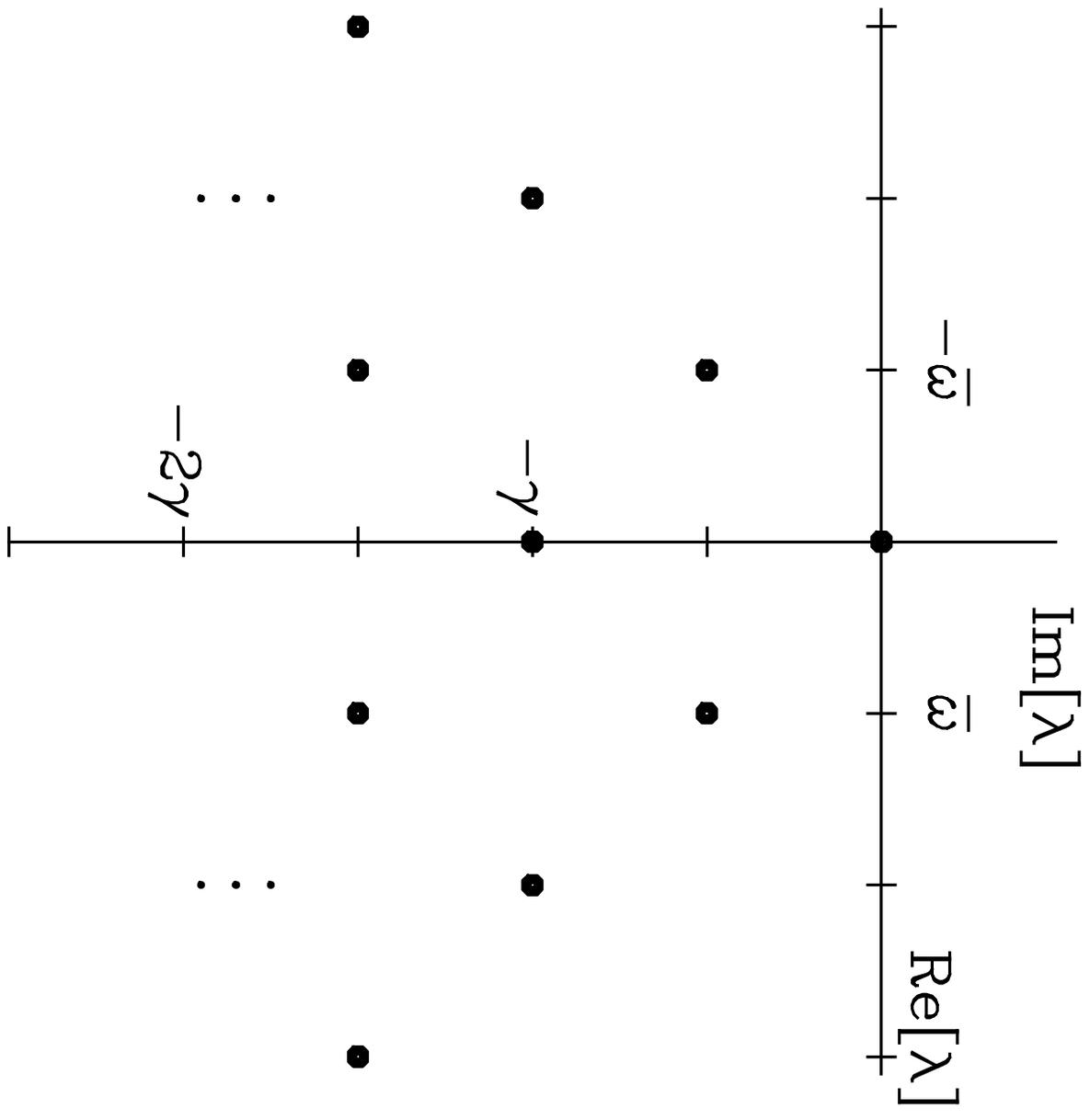
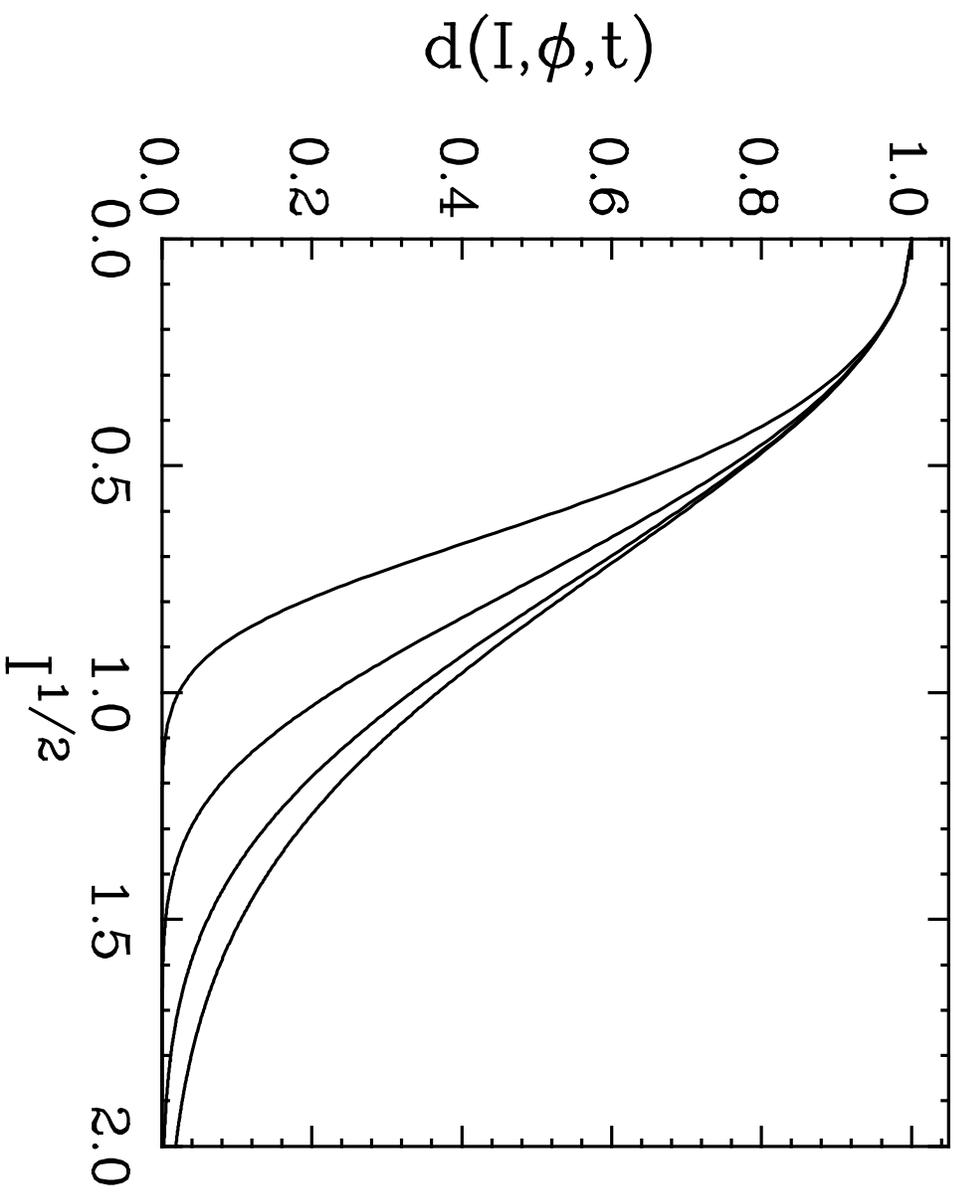
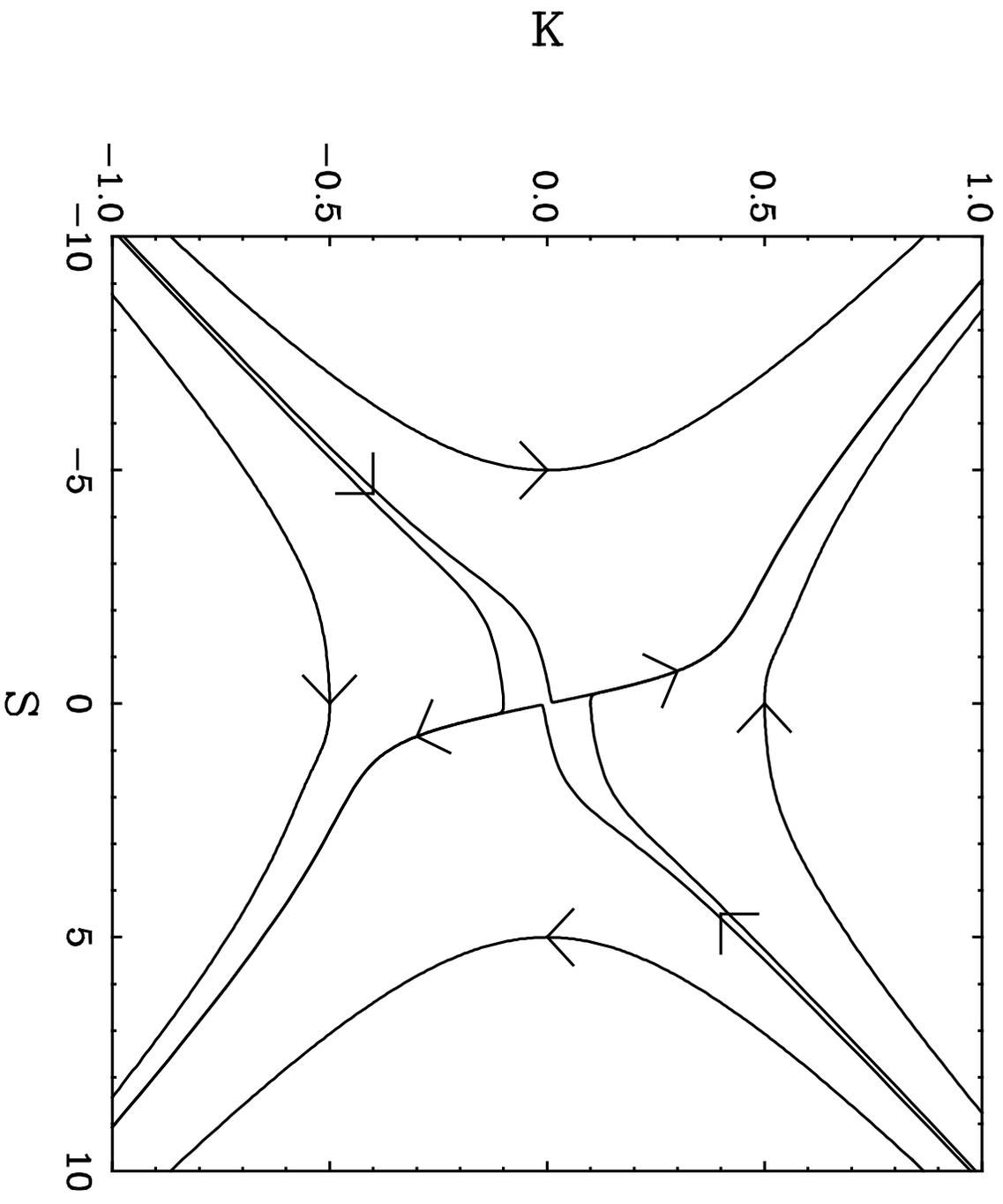
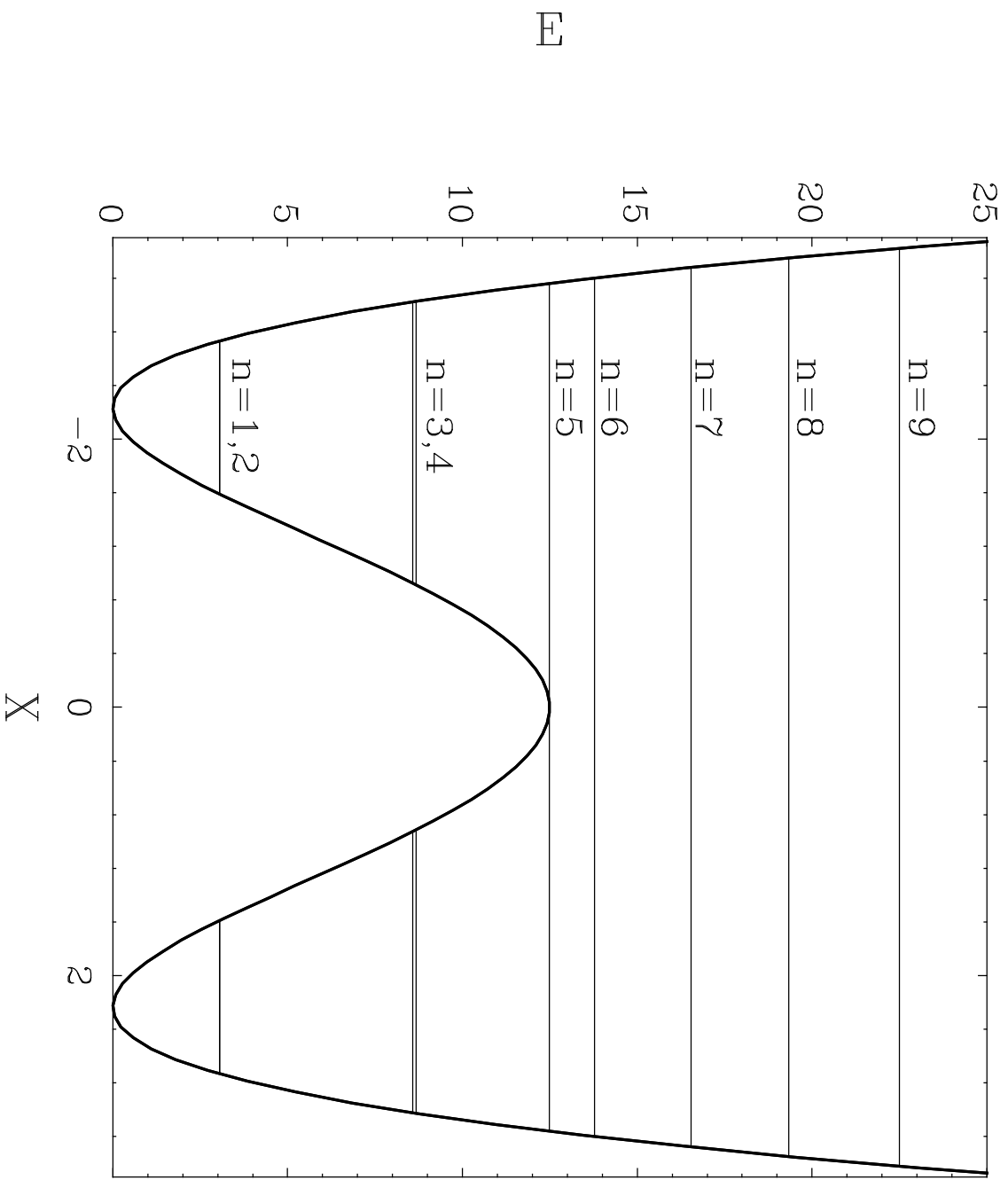
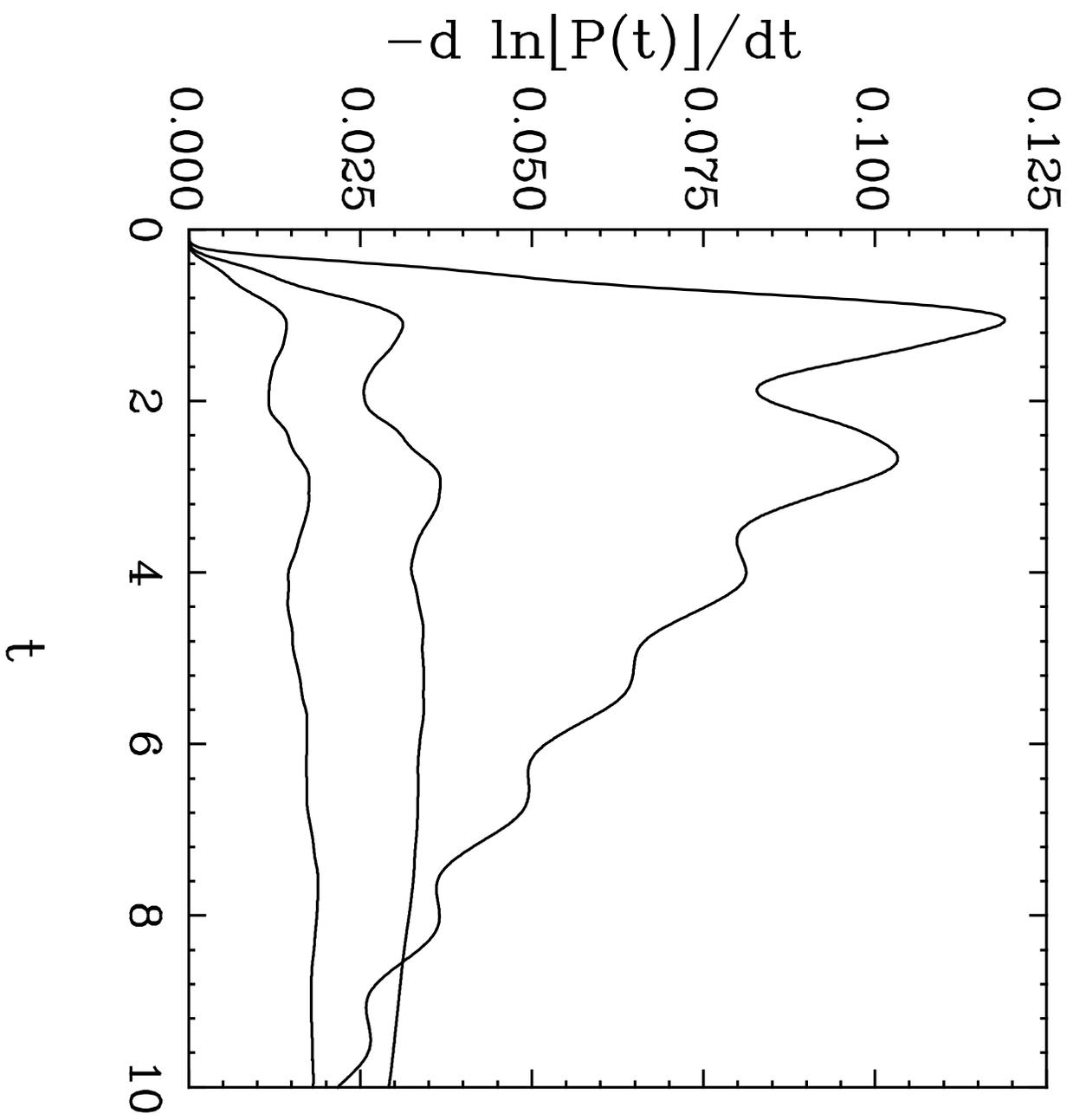
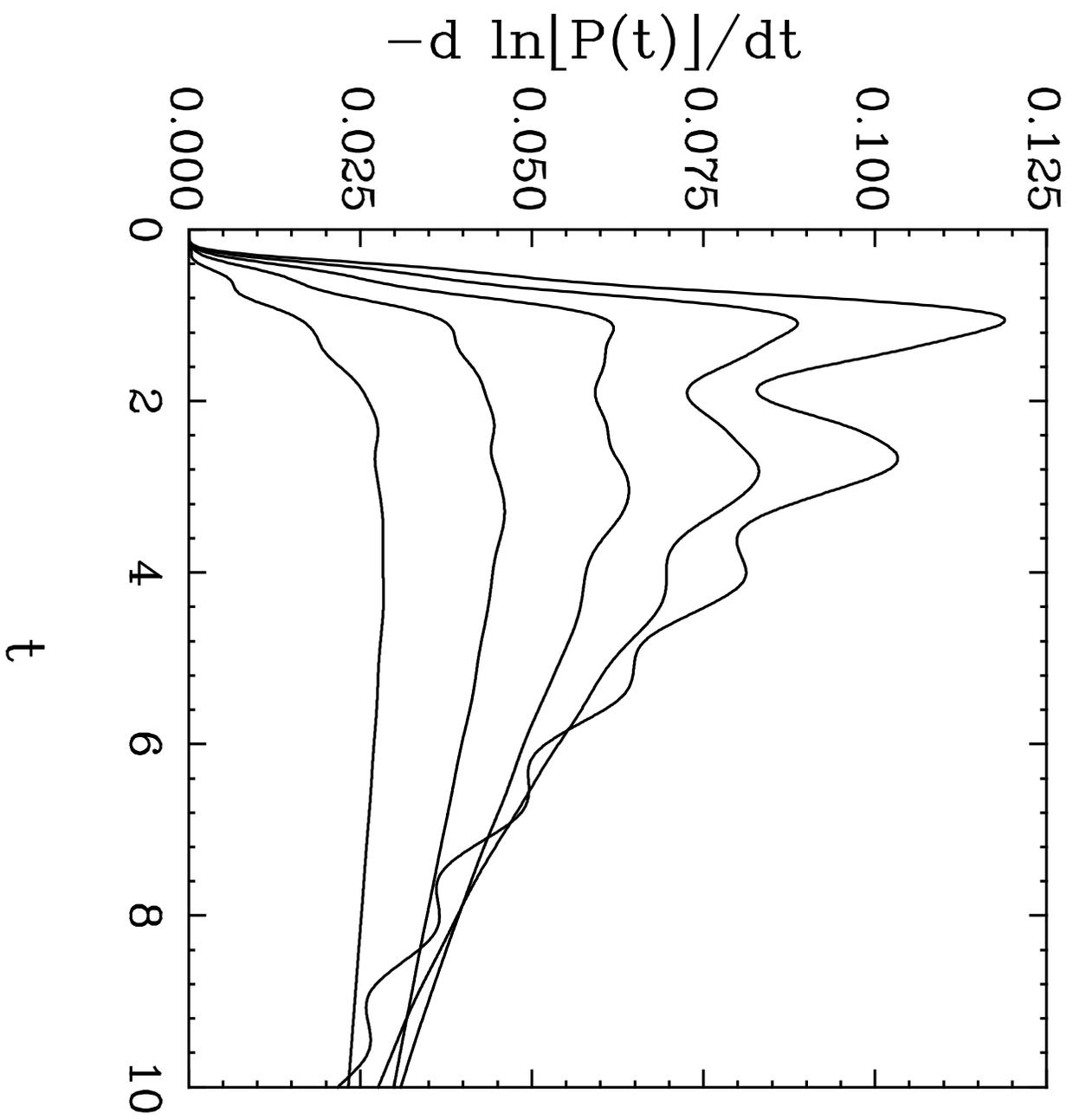
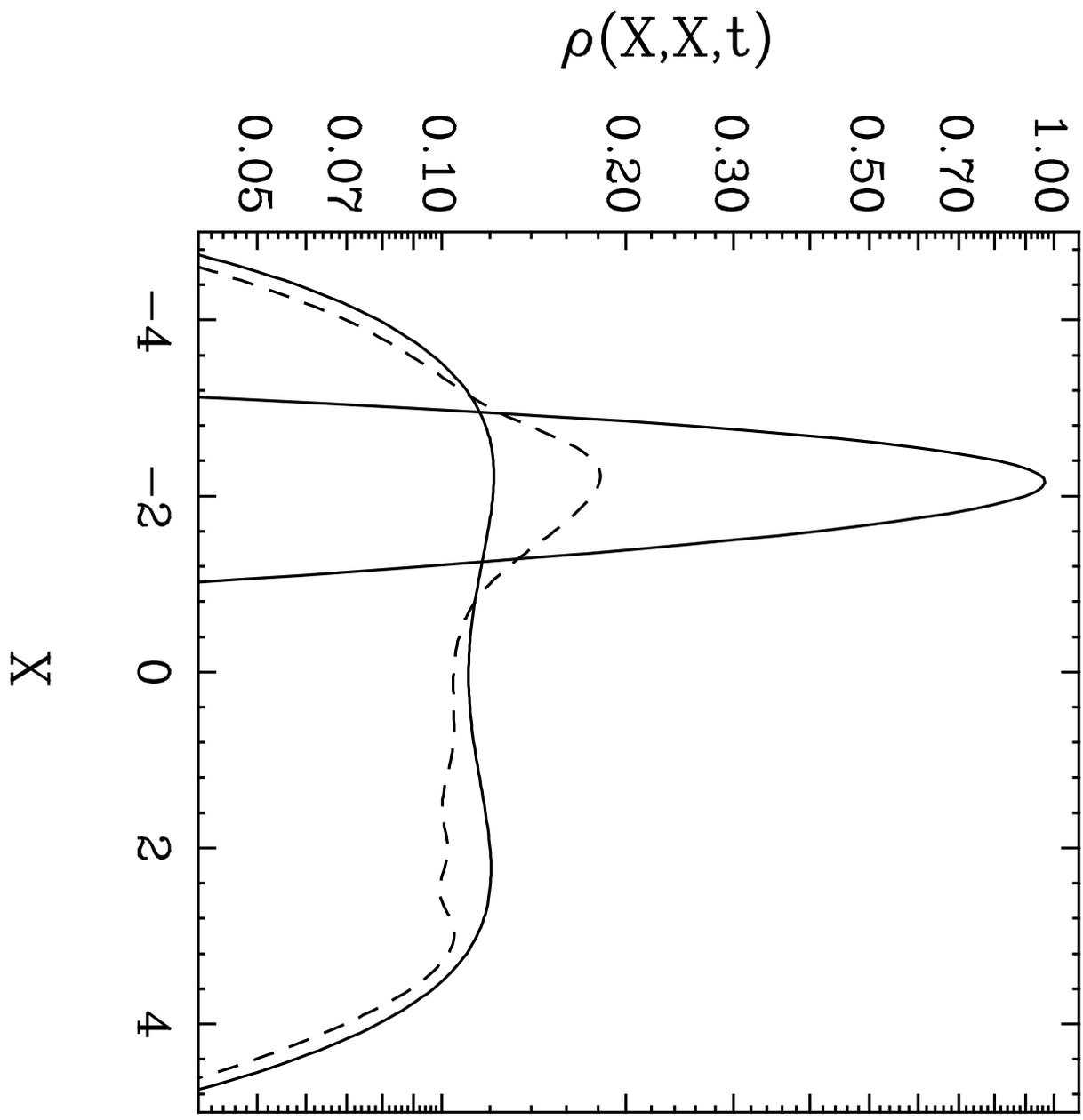
\end{document}